\documentclass[12pt,preprint]{aastex}
\usepackage{lscape}
\usepackage{longtable}
\usepackage{epstopdf}
\usepackage{graphicx}
\usepackage{epsfig}
\usepackage{natbib}
\usepackage{amsmath}

\begin{document}

\title{Searches for millisecond pulsar candidates among the unidentified \emph{Fermi} objects}

\author{
C.~Y.~Hui\altaffilmark{1},
S.~M.~Park\altaffilmark{1},
C.~P.~Hu\altaffilmark{2},
L.~C.~C.~Lin\altaffilmark{3}
K.~L. Li\altaffilmark{4}
A.~K.~H.~Kong\altaffilmark{4},
P.~H.~T.~Tam\altaffilmark{5},
J.~Takata\altaffilmark{6}, 
K.~S.~Cheng\altaffilmark{6},
Ruolan~Jin\altaffilmark{4},
T.-C.~Yen\altaffilmark{4},
Chunglee~Kim\altaffilmark{7}
}
\email{cyhui@cnu.ac.kr}
\altaffiltext{1}{Department of Astronomy and Space Science, Chungnam National University, Daejeon, Republic of Korea}
\altaffiltext{2}{Graduate Institute of Astronomy, National Central University, Jhongli 32001, Taiwan}
\altaffiltext{3}{Institute of Astronomy and Astrophysics, Academia Sinica, Taiwan}
\altaffiltext{4}{Institute of Astronomy and Department of Physics, National Tsing Hua University, Hsinchu, Taiwan}
\altaffiltext{5}{Institute of Astronomy and Space Science, Sun Yat-Sen University, Guangzhou 510275, China}
\altaffiltext{6}{Department of Physics, University of Hong Kong, Pokfulam Road, Hong Kong}
\altaffiltext{7}{Yonsei University Observatory, Yonsei University, Seoul, Republic of Korea}

\begin{abstract}
Here we report the results of searching millisecond pulsar (MSP) candidates from the \emph{Fermi} LAT second source catalog (2FGL). 
Seven unassociated $\gamma-$ray sources in this catalog are identified as promising MSP candidates based on their $\gamma$-ray properties. 
Through the X-ray analysis, we have detected possible X-ray counterparts, localized to an arcsecond accuracy.  
We have systematically estimated their X-ray fluxes and compared with the
corresponding $\gamma$-ray fluxes. The X-ray to $\gamma$-ray flux ratios for 2FGL~J1653.6-0159 and 2FGL J1946.4-5402 are comparable 
with the typical value for pulsars. For 2FGL~J1625.2-0020, 2FGL~J1653.6-0159 and 2FGL J1946.4-5402, their candidate 
X-ray counterparts are 
bright enough for performing a detailed spectral and temporal analysis to discriminate their thermal/non-thermal nature and search 
for the periodic signal. We have also searched for possible optical/IR counterparts at the X-ray positions. 
For the optical/IR source coincident with the brightest X-ray object that associated with 2FGL~J1120.0-2204, its spectral 
energy distribution is comparable with a late-type star. Evidence for the variability has also 
been found by examining its optical light curve. All the aforementioned 2FGL sources resemble a pulsar in one or more aspects, 
which make them as the promising targets for follow-up investigations.  
\end{abstract}


\section{Introduction}
Before the lanuch of \emph{Fermi} Gamma-ray Space Telescope, there was no millisecond pulsar (MSP) known as $\gamma-$ray emitter. 
Thanks to the sensitivity of the \emph{Large Area Telescope} (LAT) onboard \emph{Fermi}, a group of eight $\gamma-$ray 
MSPs has been discovered shortly after its operation commenced (Abdo et al. 2009). Since then, the population of 
$\gamma-$ray emitting MSPs continues to rise. In the second \emph{Fermi} LAT catalog of pulsars (Abdo et al. 2013), among 117 
$\gamma-$ray pulsars detected at high significance, 40 of them are MSPs and therefore form a definite class of $\gamma$-ray sources. 

In the second \emph{Fermi}-LAT source catalog (2FGL; Nolan et al. 2012), 
31\% of the $\gamma$-ray objects are unassociated with any known object. 
These unidentified \emph{Fermi} objects has formed the second largest class in 2FGL catalog. 
Among all these unidentified $\gamma$-ray objects, a number of them can possibly be MSPs. These 
$\gamma$-ray sources effectively provide a ``treasure map" for pulsation searches with radio telescopes.
As the angular resolution of LAT is much improved in comparison with its predecessors, most $\gamma$-ray sources can be localized 
to a region small enough to allow radio pulsar searches with a minimal number of telescope pointings. This strategy has been demonstrated to 
be fruitful as a large number of MSPs has been discovered through this method (see Ray et al. 2012). 

However, it should be noted that blind pulsation searches for MSPs are computational demanding. Since most of the MSPs reside in binary systems, the 
dimension of the parameter space for searching is therefore increased with the orbital parameters. Furthermore, the fast rotation of MSPs requires an 
accurate knowledge of their positions for pulsation searches. However, even with the improved angular resolution of LAT, the positional 
uncertainties of the cataloged $\gamma$-ray objects (typically a few arcmin) are still too large for constraining the MSP positions. 
Without additional information, blind pulsation searches require a fine scan of position over the error boxes of these MSP candidates. 
For $\gamma$-ray pulsation search, as very long integration time is required, the proper motion is also needed to be taken into account. 

For facilitating the MSP searches, multiwavelength observations play an important role in allevitating the aforementioned problems.
In particular, MSPs are known to be X-ray emitters. For isolated MSPs, the X-ray emission can come from the hot polar cap regions and/or the 
synchrotron radiation from the magnetospheric accelerator (e.g. Cheng \& Zhang 1999). 
Among all the newly discovered MSPs, a considerable fraction of them has shown radio eclipses (Robert 2013). These systems have very tight orbits 
with an orbital period less than a day (cf. Table~1 in Roberts 2013). Depending on the masses of their companions $M_{c}$, these systems are dubbed as 
``black widows" ($M_{c}<<0.1M_{\odot}$) or ``redbacks" ($M_{c}\gtrsim0.1M_{\odot}$) (For a recent views on these systems, please refer 
to Hui 2014). For such interacting binaries, X-ray emission can be produced through intrabinary shocks and can be orbitally modulated. 

Therefore, one can constrain
the position of the MSP candidates to an arcsecond accuracy by searching for possible X-ray counterpart within the $\gamma$-ray error box. 
If the identified X-ray sources show similar emission properties (e.g. spectral properties, $\gamma$-ray to X-ray flux ratio, variability) 
to those of known MSPs, they are considered to be promising candidates.  
Identifying possible optical counterparts with a spectrum/color similar to a late-type star/white dwarf can help us further narrow down the possible candidates. 
For the MSPs in binaries, the optical emission is originated from their companions which are heated by the relativistic wind outflow from 
the pulsars. This leads to the optical orbital modulation as seen in black widows/redbacks. If such an orbital period can be found from a candidate, 
this will make the pulsation search easier by reducing one dimension in the parameter space. The accurate optical positions can further 
help in constraining the position of MSP candidates. 

To sum up, if a candidate counterpart lies in the $\gamma$-ray positional error box of a pre-selected unidentified \emph{Fermi} object with X-ray/optical 
properties comparable with known black widow/redback systems and shows modulation with a period less than a day, it is 
very likely to be a new black widow/redback MSP. Two remarkably successful examples of MSPs identified through this scheme are 2FGL~J2339.6-0532 
(Romani \& Shaw 2011; Kong et al. 2012) and 2FGL~1311.7-3429 (Romani 2012). In this study, we present a systematic search for possible 
MSP candidates in the 2FGL catalog through a detailed X-ray/optical identification campaign. 

\section{Data Analysis and Results}
In order to identify $\gamma$-ray sources as MSP candidates, we scanned over the 2FGL catalog and selected object if it 
satisfies all the following four criteria: (1) unassociated source locates at a Galactic latitude $|b|>10^{\circ}$; (2) its variability 
index (i.e. the parameter \texttt{Variability$\_$Index} in 2FGL) is less than 41; (3) A curved spectrum (in contrast to a simple power-law) 
is required in the fitting at significance larger than $5\sigma$ (i.e. the parameter \texttt{Curve$\_$Significance} in 2FGL); 
(4) The source detection significance (i.e. the parameter \texttt{Detection$\_$Significance} in 2FGL) is larger than $10\sigma$ 
(cf. Nolan et al. 2012; Romani 2012; Ackermann et al. 2012; Kong et al. 2012). 

Condition (1) is imposed as MSPs are old objects and they should be far away from the Galactic plane which is presumably their birth places. 
For a 2FGL source with \texttt{Variability$\_$Index} $>41.64$, the probability of it to be a steady source is $<1\%$ (Nolan et al. 2012). Therefore,
we have imposed condition (2) so as to eliminate the highly variable sources such as blazars. Since the $\gamma$-ray spectra of pulsars are 
typically described by a curved model with a form of exponentially cutoff power-law (cf. Abdo et al. 2013), condition (3) can further enhance 
the chance of the selected sources as a pulsar. The last condition ensures the selected sources to be bright enough so that their positions and other 
physical parameters (e.g. $\gamma$-ray flux) have been properly constrained. 
Seven candidates are selected in accordance with these criteria. The results are summarized in Table~\ref{2fgl} which 
are arranged in the order of increasing right ascension. 

Apart from the sources in Table~\ref{2fgl}, 2FGL~J1227.1-4853, 2FGL J1311.7-3429 and 
2FGL J2339.6-0532 also fulfill our selection criteria. However, we found that these three sources have 
already identified as MSPs (Bassa et al. 2014; Romani 2012; Ray et al. 2013; Kong et al. 2012). And therefore, these three sources will not be 
considered in our study.  

\begin{figure}[t]
\centerline{\psfig{figure=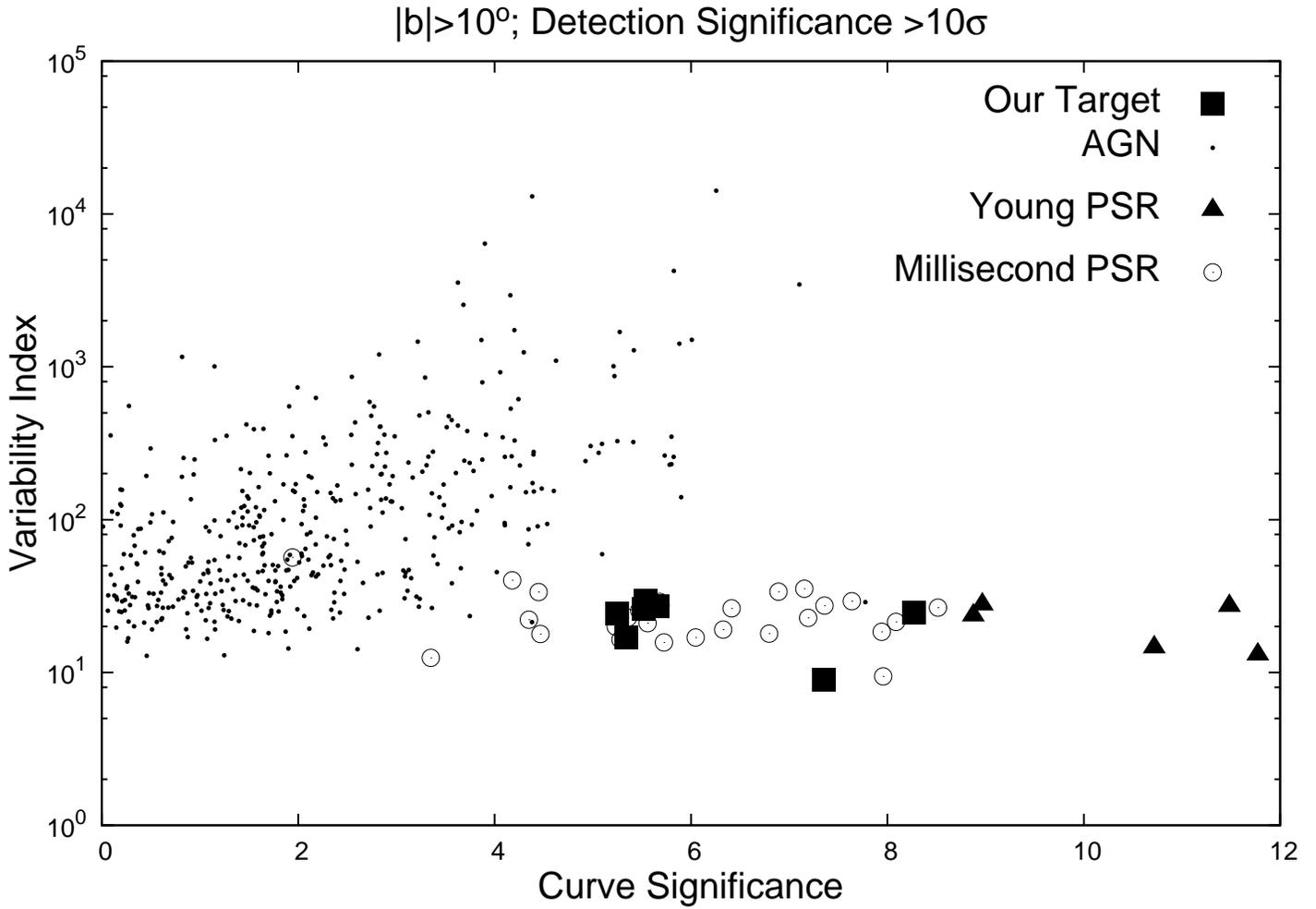,width=19cm,clip=}}
\caption[]{A plot of \texttt{Variability$\_$Index} and \texttt{Curve$\_$Significance} for all known AGNs, 
non-recycled pulsars and MSPs at $|b|>10^{\circ}$ 
which have \texttt{Detection$\_$Significance} $>10\sigma$ in 2FGL. Their distribution is compared with that of our selected targets.}
\label{c_v_plot}
\end{figure}

In Figure~\ref{c_v_plot}, we compare the distribution of \texttt{Variability$\_$Index} and \texttt{Curve$\_$Significance} of our 
selected targets with that of the known AGNs, non-recycled pulsars and MSPs at $|b|>10^{\circ}$ and have
\texttt{Detection$\_$Significance}$>10\sigma$. All the selected targets are found in a region occupied by the MSPs which justifies 
our adopted selection criteria. For the 30 known MSPs in this region, only 4 of them are isolated. This demonstrates our 
adopted scheme can be effective in spotting MSPs in binaries. On the other hand, 5 young/non-recycled pulsars 
and an AGN are also found in this region. 

\subsection{Analysis of the X-ray point sources in the $\gamma-$ray error ellipses}

To search for the possible X-ray counterparts associated with our short-listed 2FGL sources, we utilized archival X-ray spectral 
imaging data and looked for all the X-ray sources which lie within their $\gamma$-ray positional uncertainties. 
Very recently, the third {\it Fermi} $\gamma-$ray point sources catalog (3FGL) has been released (Acero et al. 2015). 
With four years LAT data, 3FGL catalog can possibly provide an improved localization of the $\gamma-$ray sources. In searching 
for possible X-ray counterparts, we focused on X-ray point-like sources detected within the smallest possible 95\% $\gamma-$ray confidence error 
ellipses provide by 2FGL/3FGL.
The results are summarized 
in Table~2. All these X-ray sources are detected at a significance larger than 
$4\sigma$. Their observed and absorption-corrected fluxes are systematically computed with the aid of \emph{PIMMS} (ver. 4.7) 
by assuming an absorbed power-law spectrum with a 
photon index $\Gamma=2$ and a column absorption consistent with the Galactic HI column density in the corresponding direction 
(Kalberia et al. 2005). The details of the X-ray analyses are discussed in the followings.

\subsubsection{2FGL J1120.0-2204}
\begin{figure}[t]
\centerline{\psfig{figure=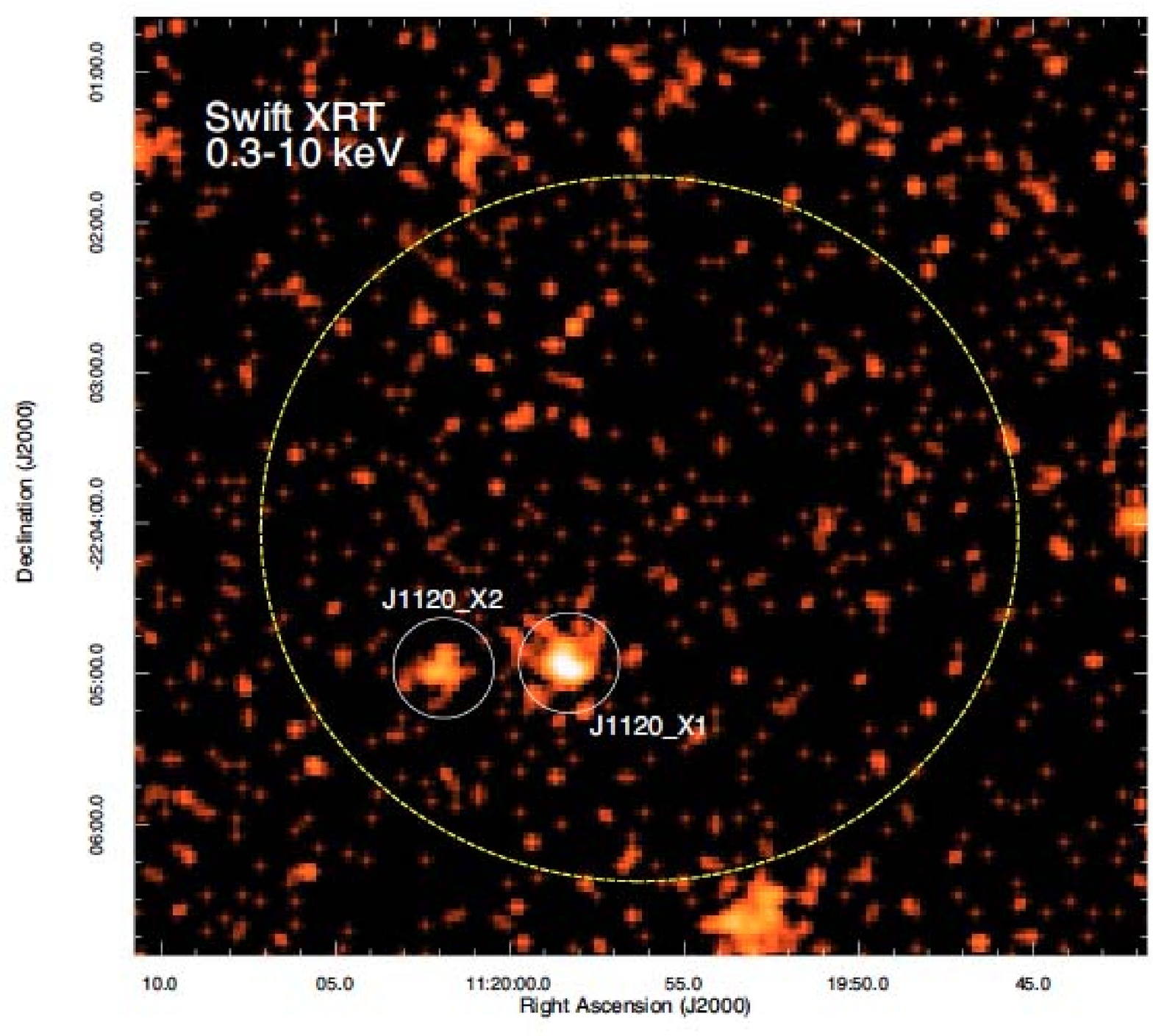,width=19cm,clip=}}
\caption[]{The potential X-ray counterparts (white circles) of 2FGL~J1120.0-2204 as observed by \emph{Swift} XRT. This image is produced by 
using all 
the available XRT data of this field. 
The dashed ellipse illustrates the 3FGL 95\% confidence ellipse. The image is smoothed with a Gaussian kernal of $\sigma=4''$.}
\label{j1120_x_img}
\end{figure}

2FGL~J1120.0-2204 have been observed by \emph{Swift} XRT with multiple snapshots. 
With all the available \emph{Swift} XRT data combined, we have an effective exposure of $\sim65$~ks 
in this field. Utilizing a wavelet source detection algorithm, we have searched for its possible X-ray counterparts. 
We binned the whole data into a $1024\times1024$ image and 
we set the detection threshold such that no more than one false detection caused by background
fluctuation is in the whole field. 
Two sources, J1120$\_$X1 and J1120$\_$X2, with a signal-to-noice 
ratio $>4\sigma$ were found in its 3FGL 95\% $\gamma-$ray error ellipse (see Figure~\ref{j1120_x_img}). Their basic properties, including 
X-ray fluxes and the statistical position uncertainties, are summarized in Table~2. The X-ray fluxes were 
calculated by using the total Galactic HI column density, $3.9\times10^{20}$~cm$^{-2}$, in this direction (Kalberla et al. 2005) 
and assuming a power-law model with $\Gamma=2$. 

We have also computed the probability for one or more X-ray sources lying in the $\gamma$-ray error ellipse by chance.  
We counted the number of X-ray sources detected in the whole FoV and computed the source density. Based on this, we estimated 
the number of chance coincidences $\lambda$ expected within the $\gamma$-ray error ellipse. Assuming a Poisson distribution, 
the probability of finding one or more chance coincidences is given by:

\begin{equation}
P\left(n\geq 1\right)=\sum_{n=1}^{\infty}\frac{\lambda^{n}e^{-\lambda}}{n!}=1-e^{-\lambda}
\end{equation}

\noindent For 2FGL~J1120.0-2204, we found that $P\left(n\geq 1\right)\sim56\%$. 

The limiting flux for a $4\sigma$ point source detection in this field is $3.7\times10^{-15}$~erg~cm$^{-2}$~s$^{-1}$ 
in an energy range of 0.3-10~keV. 
Apart from the basic source characterization presented in Table~2, we have also performed detailed temporal and spectral analysis 
for those sources with more than 100 net counts detected throughout our investigation. 
Since the net counts of J1120$\_$X1 and J1120$\_$X2 are below 100, no further X-ray analysis will be conveyed.

\subsubsection{2FGL J1539.2-3325}
By merging all the available data from \emph{Swift} XRT observations in the field of 2FGL~J1539.2-3325, 
we have an effective exposure of $\sim84$~ks. We have attempted to search for possible X-ray counterparts within 
its 3FGL 95\% $\gamma-$ray error ellipse. However, we cannot detect any source with a significance $>4\sigma$ with the current data.
We place a limiting flux of $6.1\times10^{-15}$~erg~cm$^{-2}$~s$^{-1}$ for the $4\sigma$ point source detection in this field.

\subsubsection{2FGL J1625.2-0020}
2FGL~J1625.2-0020 has been observed by \emph{XMM-Newton} on 8 February 2012 for a total exposure of $\sim26.6$~ks 
(Obs. ID: 0672990401) with both MOS cameras operated in full-frame mode and PN camera operated in extended full-frame mode.
Thin filters have been used to minimize the optical contamination for all cameras throughout the observation. 
With the most updated instrumental calibration, we
generated the event lists from the raw data obtained from all EPIC instruments
with the tasks \emph{emproc} and \emph{epproc} of the {\bf XMM}-Newton {\bf S}cience {\bf A}nalysis
{\bf S}oftware (XMMSAS version 12.0.1). The event files were subsequently filtered for the energy range
from 0.3 keV to 10 keV and selected only those events for which the pattern was
between $0-12$ for MOS cameras and $0-4$ for the PN camera.
We further cleaned the data by accepting only the good times when sky background was low in the whole CCD and
removed all events potentially contaminated by bad pixels. 
After the filtering, 
the effective exposures are found to be 14.9~ks, 14.0~ks and 10.7~ks for MOS1, MOS2 and
PN respectively. All the subsequent analyses were performed in the energy range of $0.3-10$~keV. 
 
\begin{figure}[t]
\centerline{\psfig{figure=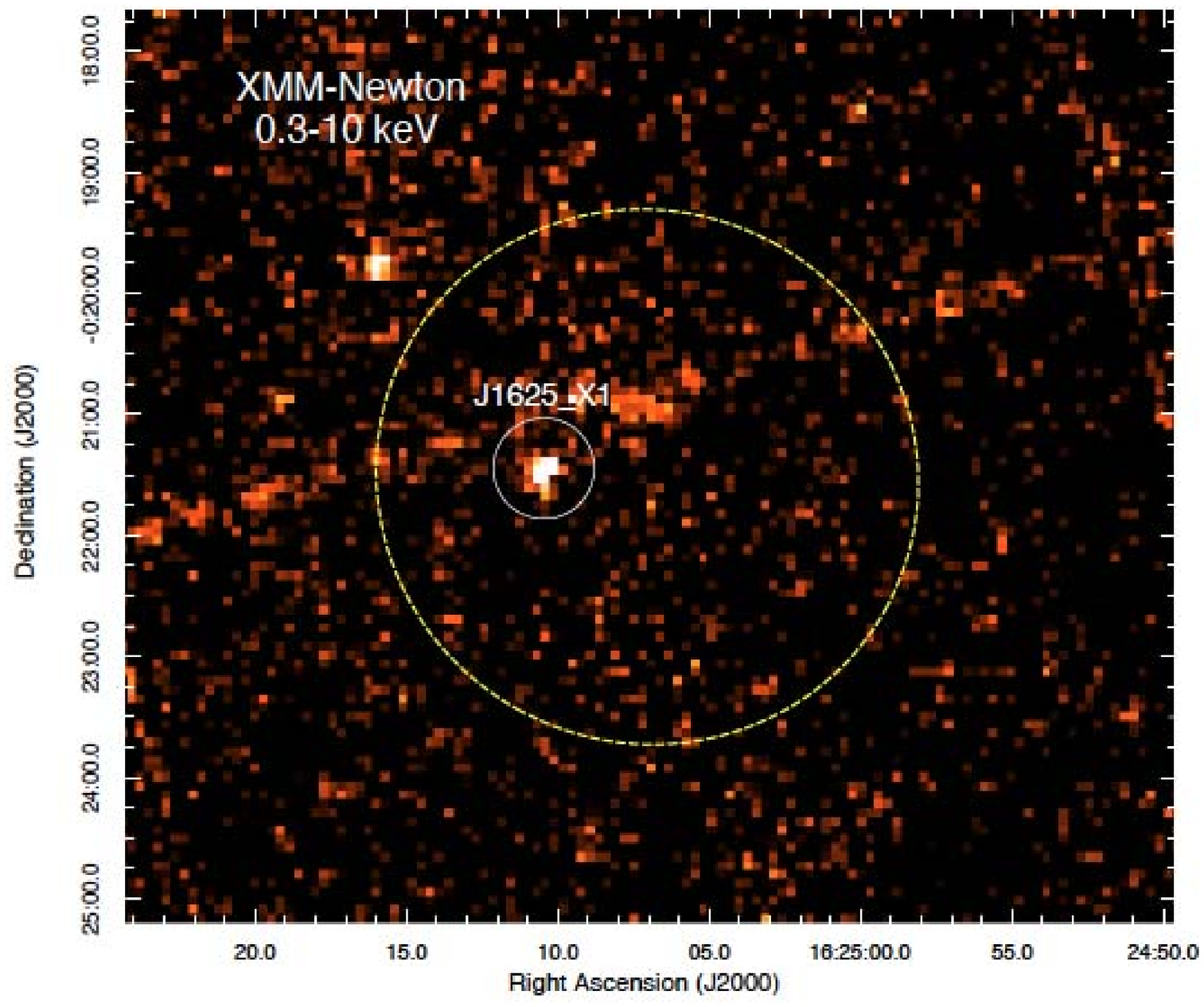,width=19cm,clip=}}
\caption[]{The potential X-ray counterpart (white circle) of 2FGL~J1625.2-0020 as observed by \emph{XMM-Newton}. 
This image is produced by merging all three CCD data.
The dashed ellipse illustrates the 3FGL 95\% confidence ellipse. 
The image is smoothed with a Gaussian kernal of $\sigma=4''$.}
\label{j1625_x_img}
\end{figure}

We performed a source detection by using maximum likelihood fitting on MOS1, MOS2 and PN data individually 
with the aid of the XMMSAS task
\emph{edetect\_chain}. The detection threshold was chosen to be $4\sigma$. 
Source lists resulted from three individual cameras are subsequently
correlated and merged by using the XMMSAS task \emph{srcmatch}. 
In case the position of a source obtained from two detections are consistent 
within their $3\sigma$ uncertainties, they are merged as a single entry. 
 We have detected only one source, J1625$\_$X1, 
within the 3FGL 95\% $\gamma$-ray error ellipse of 2FGL~J1625.2-0020 (see Figure~\ref{j1625_x_img}).
The limiting flux for $4\sigma$ source detection is 
$1.1\times10^{-14}$~erg~cm$^{-2}$~s$^{-1}$  (0.3-10 keV).
Following the procedure described in \S2.1.1 and Equation (1), the probability of one or more chance coincidences lying 
within the $\gamma$-ray error ellipse is $\sim50\%$. 

Since J1625$\_$X1 has a net counts of $\sim120$~cts, 
we have carried out a detailed analysis for it. 
We extracted the X-ray spectrum of J1625$\_$X1 from a circular region with 
a radius of 20~arcsec centered at its position given in Table~2.  
The background spectrum was  sampled from a nearby source-free region in the individual camera. 
The response files were produced by the XMMSAS task \emph{rmfgen} and \emph{arfgen}. 
The spectrum was binned so as to have $>10$ counts per spectral bin. 

We firstly examined the spectrum with 
an absorbed power-law model. In view of the limited photon statistics, we fixed the column absorption at 
$N_{H}=5.8\times10^{20}$~cm$^{-2}$ as inferred by the HI column density (Kalberla et al. 2005) and adopted 
$C$ statistic for the modeling fitting (Cash 1979). 
Although the power-law model can provide a statistical acceptable fit ($C=9.90$ for 14 d.o.f.), 
it results in an unphysically large photon index $\Gamma=3.1^{+0.4}_{-0.3}$. 
Therefore, the non-thermal scenario is not favored. 
All the quoted uncertainties of the spectral
parameters are $1\sigma$ for 1 parameter of interest.

We have also examined its spectrum with an absorbed blackbody model. With $N_{H}$ fixed at 
$5.8\times10^{20}$~cm$^{-2}$ it yields a temperature of $kT=0.16\pm0.02$~keV and an emission radius 
of $R=162^{+64}_{-47}d_{\rm kpc}$~m with 
$C=12.50$ for 14 d.o.f., where $d_{\rm kpc}$ is source distance in unit of 1~kpc. 
Since this model provides a more reasonable description of the data than the power-law, 
we conclude that the X-ray emission from J1625$\_$X1 has a thermal origin. 

One important indication for a black widow/redback MSP is the presence of
X-ray orbital modulation which results from intrabinary shock (e.g. Tam et al. 2010; Huang et al. 2012; Hui et al. 2014).
For the temporal analysis, we have extracted the light curve of J1625$\_$X1 and subtracted the background by adopting the regions used in the spectral analysis. 
All its arrival times were firstly barycentric-corrected by using its X-ray position reported in Table~2. 
Updated planetary ephemeris JPL DE405 is used throughout this investigation.
We have searched for periodic 
signals to $\sim1.5$~hrs which corresponds to the half of the effective exposure. We did not find any 
promising periodicity in this analysis. 

For testing the robustness of the results, we have repeated the analysis with various background sampled from the 
nearby source-free regions. We found the results obtained from these independent analyses are consistent with each other. 

\subsubsection{2FGL J1653.6-0159}
2FGL J1653.6-0159 has been observed by \emph{Chandra} ACIS-I on 2010 Jan 24 with an exposure of $\sim21$~ks (Obs. ID: 11787). A previous study has 
reported the X-ray analysis of this data (Cheung et al. 2012). However, as a number of investigations which are crucial for 
identifying the possible MSP nature, such as a detailed spectral analysis and timing analysis, was not reported in Cheung et al. (2012), we decided 
to re-analyse this data set. 

By using the script \emph{chandra\_repro} provided
in the {\bf C}handra {\bf I}nteractive {\bf A}nalysis {\bf O}bservation software (CIAO~4.6), we have reprocessed the data
with CALDB (ver. 4.6.1.1). In order to utilize the superior angular resolution of \emph{Chandra} to tightly constrain the X-ray positions, 
we have applied sub-pixel event repositioning during the data reprocessing in order
to improve the positional accuracy of each event (cf. Li et al. 2004). We restricted the analysis of
this ACIS data in an energy range of $0.5-8$~keV.

\begin{figure}[t]
\centerline{\psfig{figure=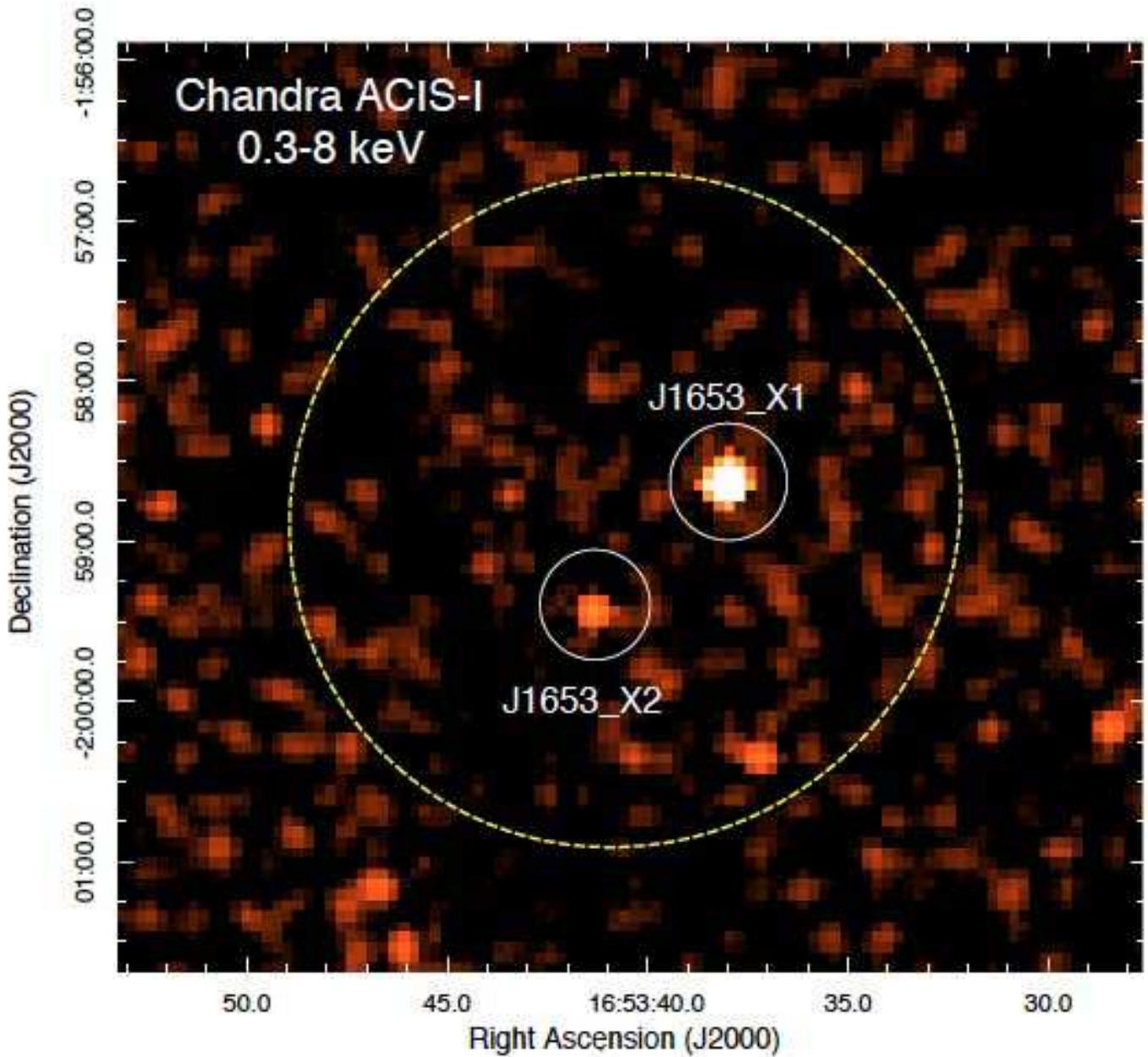,width=19cm,clip=}}
\caption[]{The potential X-ray counterparts (white circles) of 2FGL~J1653.6-0159 as observed by \emph{Chandra}. The dashed ellipse illustrates the 
95\% 3FGL confidence ellipse. The image is smoothed with a Gaussian kernal of $\sigma=5''$. }
\label{j1653_x_img}
\end{figure}

By means of the wavelet source detection algorithm (CIAO tool: {\it wavdetect}),
we searched for possible X-ray counterparts of 2FGL~J1653.6-0159. The exposure variation across the detector was
accounted by the exposure map. 
Within the 3FGL 95\% error ellipse of 2FGL~J1653.6-0159, only two X-ray sources can be detected at a significance 
$>4\sigma$ (cf. Table~2 and Figure~\ref{j1653_x_img}). 
The limiting flux for a $4\sigma$ detection in this field is $1.4\times10^{-14}$~erg~cm$^{-2}$~s$^{-1}$ 
(0.3-10~keV). The probability of one or more chance coincidences lying
within the $\gamma$-ray error ellipse is $\sim63\%$. 

Since J1653$\_$X1 has $\sim360$ net counts collected 
with a circular aperture with a radius of 2~arcsec, 
we therefore proceeded to examine its X-ray properties in further details. 
Very recently, 
a possible orbital modulation with a period of $P=4488$~s was independently reported by Romani et al. (2014) and Kong et al. (2014) using optical observations.
This information can enable us to investigate the X-ray orbital modulation. 
Before we examined the temporal behavior of J1653$\_$X1,
its arrival times were firstly barycentric-corrected by using its X-ray position reported in Table~2. 
For the background subtraction, we have adopted an annular region with inner/outer radius of 5.5~arcsec/10~arcsec 
centered at its nominal X-ray position. The background-subtracted light curve folded at $P=4488$~s is shown in Figure~\ref{j1653_x1_lc}. $\chi^{2}$ test 
indicates that the distribution in Figure~\ref{j1653_x1_lc} differs from a uniform distribution at a confidence level of $\sim99.2\%$. 

In order to probe its emission nature in 
details, we have also re-examined the X-ray spectrum of J1653$\_$X1. The source spectrum and the background spectrum were sampled from the same 
regions adopted for the aforementioned temporal analysis. The spectrum was grouped to have at least 15 counts per spectral bin. We found that 
the X-ray spectrum of J1653$\_$X1 can be described by an absorbed power law model ($C$ statistic=15.50 for 19 d.o.f.). 
The best-fit model 
yields a column density of $N_{H}=9.0^{+11.6}_{-9.0}\times10^{20}$~cm$^{-2}$,
a photon index of $\Gamma=1.6^{+0.2}_{-0.1}$ and a
normalization of $3.6^{+0.9}_{-0.7}\times10^{-5}$~photons~keV$^{-1}$~cm$^{-2}$~s$^{-1}$ at 1 keV. 
We noted that the inferred $N_{H}$ is comparable with a total Galactic HI column 
density of $8.2\times10^{20}$~cm$^{-2}$ (Kalberla et al. 2005). 
On the other hand, we found that a blackbody is not able to provide any reasonable description of the data 
($C$ statistic=57.60 for 19 d.o.f.). 
Therefore, we concluded that the X-ray emission from J1653$\_$X1 is mostly non-thermal dominant. 

Although a simple power-law can provide a reasonable overall description of the data, we have identified systematic deviations at the energies around 
3.5~keV by examining the fitting residuals (see Figure~\ref{j1653_spec}). In examining the Figure~5 in Cheung et al. (2012), there was also an indication for the residuals around this energy.
We speculated that this might indicate the presence of an emission line feature. To test this 
hypothesis, we have added an additional gaussian component to the power-law model for the spectral fitting. 
This yielded a best-fit gaussian 
line profile at energy $E=3.4\pm0.1$~keV with a width of $\sigma=0.17^{+0.12}_{-0.17}$~keV and the power-law with 
$\Gamma=1.7\pm0.2$ with
a normalization of $3.6^{+1.0}_{-0.7}\times10^{-5}$~photons~keV$^{-1}$~cm$^{-2}$~s$^{-1}$ at 1 keV. The best-fit column absorption 
$N_{H}=9.0^{+11.9}_{-9.0}\times10^{20}$~cm$^{-2}$ is consistent with the simple power-law fit model and the HI column density within the 
tolerance of statistical uncertainties. The corresponding goodness-of-fit is $C=10.18$ for 16 d.o.f..

To further examine the evidence for the line feature, we have simulated the null distribution of likelihood ratio between the alternative model 
(i.e. absorbed power-law plus gaussian) and the null model (i.e. absorbed power-law model) following the method suggested by Protassov et al. (2002). 
With the aid 
of \emph{Sherpa}, we simulated 10000 data sets with Poisson noise according to the best-fit parameters of the null model. The simulation was 
performed with the same instrumental responses and the exposure of the observed data. Each of the simulated data was fitted with the null and 
alternative model. The resultant distribution of the likelihood ratio implies 
a $p$-value of $\sim11\%$. Therefore, based on this test, there is no solid evidence for the emission line. A deeper observation is 
encouraged for investigating this feature. 

\subsubsection{2FGL J1729.5-0854}
For 2FGL~J1729.5-0854, we notice that its 95\% $\gamma-$ray error ellipse in 3FGL ($0.58^{\circ}\times0.45^{\circ}$) is 
larger than that in 2FGL ($0.20^{\circ}\times0.18^{\circ}$). To investigate this, we have performed a brief $\gamma-$ray analysis 
of this source by using the $\sim6$~years LAT data. We subtracted the background contribution by including the 
Galactic diffuse model (gll\_iem\_v05\_rev1.fits) and the isotropic background (iso\_source\_v05.txt), 
as well as all sources in the 3FGL catalog within the circular region of 25$^\circ$ radius around 2FGL~J1729.5-0854.
In Figure~\ref{j1729_lat_img}, we show the background-subtracted count map
at energies $>100$~MeV. 
2FGL~J1729.5-0854 is apparently extended. If the extended feature is genuine and it is indeed associated 
with a pulsar, the feature might be originated from a pulsar wind nebula. However, given the current source significance, it is 
difficult to discern whether 2FGL~J1729.5-0854 is truely extended or it consists of more than one unresolved point source. 
Furthermore, the inadequacy of the adopted diffuse model in a possibly complex region can also lead to a poor localization 
accuracy of the source. 
A further dedicated $\gamma$-ray analysis of this source is encouraged to identify its nature. 

With the above caveats in mind, to search for possible X-ray counterparts we considered the error ellipse given by 2FGL.
By merging all the available data from \emph{Swift} XRT observations in the field of 2FGL~J1729.5-0854,  
we have an effective exposure of $\sim57$~ks. We have detected 13 X-ray sources within its 2FGL 95\% $\gamma-$ray error ellipse 
(see Figure~\ref{j1729_x_img}). Their X-ray properties are summarized in Table~2.
The probability of one or more chance coincidences lying
within the $\gamma$-ray error ellipse is $\sim99.9\%$. This indicates a large fraction 
of the detected sources might lie in the $\gamma-$ray error ellipse by chance, which mainly due to the relatively large $\gamma$-ray 
positional error. 
The limiting flux for a $4\sigma$ point source detection in this field is $9.3\times10^{-15}$~erg~cm$^{-2}$~s$^{-1}$
in an energy range of 0.3-10~keV.

\subsubsection{2FGL J1744.1-7620}
2FGL~J1744.1-7620 has been observed by \emph{XMM-Newton} on 21 August 2012 for a total exposure of $\sim25.9$~ks (Obs. ID: 0692830101). 
While both MOS1 and MOS2 cameras were operated in full-frame mode with a medium filter, 
PN camera was operated in extended full-frame mode 
with a thin filter. 
For the calibration and data reduction, we adopted the procedures as we described in \S2.1.3. After applying the 
good-time-interval filtering by accepting the times when the sky background was low in the CCD and removing all events
potentially contaminated by bad pixels, the effective
exposures are found to be 25.3~ks, 25.3~ks and 17.0~ks for MOS1, MOS2 and PN, respectively.

Only one source, J1744$\_$X1, is detected within the 
3FGL 95\% $\gamma$-ray error ellipse
of 2FGL~J1744.1-7620 at a signal-to-noise
ratio $>4\sigma$ by using the XMMSAS task \emph{edetect\_chain} and \emph{srcmatch} (see Figure~\ref{j1744_x_img}). 
Its basic X-ray properties are summarized in Table~2. 
The probability of one or more chance coincidences lying
within the $\gamma$-ray error ellipse is $\sim48\%$.
The limiting flux for a $4\sigma$ detection in this field is $8.3\times10^{-15}$~erg~cm$^{-2}$~s$^{-1}$
(0.3-10~keV). 

Besides this \emph{XMM-Newton} observation, there was a $\sim42$~ks \emph{Suzaku} observation of 2FGL~J1744.1-7620 performed 
on 14 April 2010 (Obs. ID 705013010). However, we found that J1744$\_$X1 cannot be detected in this data. 
Adopting the total Galactic HI column density of $8.5\times10^{20}$~cm$^{-2}$ (Karbela et al. 2005) and 
a power-law model with $\Gamma=2$, we place a 
$3\sigma$ limiting  of $6.3\times10^{-14}$ 
erg~cm$^{-2}$~s$^{-1}$ in 0.3-10~keV which is above the observed flux of 
J1744$\_$X1 in the same energy range (see Table~2). Therefore, the null-detection of this \emph{Suzaku} observation 
can be ascribed to the poor instrumental sensitivity for point sources. 

\subsubsection{2FGL J1946.4-5402}
2FGL J1946.4-5402 was observed by { \it Suzaku} on 31 October 2011 with an exposure of $\sim$42 ks (Obs. ID 706026010).  
Figure~\ref{j1946_x_img} shows the image extracted from the merged observations obtained with the two front-illuminated detectors (XIS0 and XIS 3) 
and one back-illuminated detector (XIS1). 
Within the 3FGL 95\% error ellipse of 2FGL J1946.4-5402, only one source, J1946$\_$X1, can be detected with a signal-to-noise ratio $\sim17\sigma$. 
The probability of one or more chance coincidences lying
within the $\gamma$-ray error ellipse is $\sim24\%$. 
The limiting flux of $4\sigma$ source detection is $3\times10^{-14}$~erg~cm$^{-2}$~s$^{-1}$ (0.3-10~keV). 

Its nominal position is  RA(J2000)=$19^{\rm h}46^{\rm m}34.241^{\rm s}$ and Dec(J2000)=$-54^{\circ}02'32.96''$.  
Nevertheless, due to its inferior spatial resolution, the positional uncertainty resulted from {\it Suzaku} observation is typically at an 
order of $\sim1'$ (Uchiyama et al. 2008). Therefore, it is not possible to constrain the position of J1946$\_$X1 to an arcsecond accuracy 
with the \emph{Suzaku} data. 

In order to better constrain the X-ray position, we have observed 2FGL~J1946.4-5402 with \emph{Swift} XRT on 14 November 2014 
(Obs. ID 00033525001). With a 6.4~ks exposure, $\sim20$ photons from J1946$\_$X1 were detected in $0.3-10$~keV. With an improved positional 
determination by \emph{Swift} XRT, 
its X-ray position is constrained to be RA(J2000)=$19^{\rm h}46^{\rm m}33.694^{\rm s}$ and Dec(J2000)=$-54^{\circ}02'34.91''$ with a 
positional uncertainty of 3.9" (see Table~2). 

For both spectral and temporal analysis, we used the \emph{Suzaku} data which provides a desirable photon statistic. 
The source events within a circular region with a radius of 1.5 arcmin around J1946$\_$X1 were selected, which corresponds to a $\sim 75\%$ encircled 
energy fraction. The background events were sampled from a nearby source-free region. After background subtraction, there are $\sim153$~cts, $\sim215$~cts and 
$\sim161$~cts available from XIS0, XIS1, and XIS3, respectively. 
For the spectral analysis, the response files were generated with the latest Suzaku/XIS calibration files 
(20140701). The spectra were grouped so as to have at least 20 counts per spectral bin. All the spectral fits were performed in the energy range 
of $0.2-10$~keV. 

In examining the observed spectra with an absorbed blackbody model, we found that it cannot yield any statistically reasonable fit (with 
$\chi^{2}=45.63$ for 18 d.o.f.). On the other hand, an absorbed power-law model can describe the data fairly well ($\chi^{2}=20.70$ for 18 d.o.f.) 
which clearly indicates the X-ray emission from J1946$\_$X1 is dominated by non-thermal emission. 
However, in estimating the uncertainty of the column absorption, we found that this parameter cannot be properly constrained. Therefore, 
in all subsequent analysis, we fixed 
$N_{H}$ at $4.3\times10^{20}$~cm$^{-2}$ as inferred from the total Galactic HI column density (Kalberla et al. 2005). The power-law
fit with $N_{H}$ fixed at this value yields $\Gamma=1.5\pm0.2$ and a normalization of $(1.4\pm0.4)\times10^{-5}$~photons~keV$^{-1}$~cm$^{-2}$~s$^{-1}$ 
at 1 keV. The corresponding goodness-of-fit is $\chi^{2}=22.35$ for 19 d.o.f. which provides an acceptable description of the observed data.

We have performed a periodicity search for J1946$\_$X1. After barycentric correcting all the events, 
we first applied the epoch-folding method to look for possible signals.  
However, no significant signal was detected except for the one related to the orbital period of {\it Suzaku} (96 min).  
To avoid the possible contamination caused by background, we also worked on background subtracted light curve and 
searched for periodicity by using the Lomb-Scargle periodogram (Scargle 1982; Press \& Rybicki 1989) 
and the analysis of variance (Schwarzenberg-Czerny 1982). No significant periodic signal was detected.

\subsection{Searches for optical/IR counterparts}
With the aforementioned X-ray analysis, we have constrained the positions of the potential X-ray 
counterparts possibly associated with the $\gamma$-ray MSP candidates 
to an arcsecond accuracy. Such well-determined positions enable us to further look for their possible optical/IR counterparts.
For a MSP in a binary, the optical/IR emission is presumably originated from its companion.  
Therefore, this search can provide insight on the nature of their companions.  
Utilizing USNO-B1.0 catalog (Monet et al. 2003), we firstly searched for possible optical counterparts. 
In order to minimize the chance of misidentification, we selected the optical sources with their 
proper-motion corrected positions lie within the X-ray positional error of each source by combining their 
statistical uncertainties in Table~2 and the corresponding systematics affecting absolute astrometry in quadrature.
\footnote{{\it Swift}: http://heasarc.nasa.gov/docs/heasarc/caldb/swift/docs/xrt/SWIFT-XRT-CALDB-07$\_$v4.pdf}
\footnote{{\it XMM-Newton}: http://xmm2.esac.esa.int/docs/documents/CAL-TN-0018.pdf }
\footnote{{\it Chandra}: http://cxc.harvard.edu/cal/ASPECT/celmon/}
6 potential counterparts have been identified with this selection criterion. We noted that the position of the 
possible optical counterpart of J1653$\_$X1 differs from its X-ray position by 1.14 arcsec 
which fails to meet our predefined criterion. However, through a dedicated optical temporal investigation, 
Kong et al. (2014) have revealed the optical emission modulated at the same periodicity as X-ray and therefore 
optical identification is secured. For completeness, we have included this source and the 
results are summarized in Table~3. 
Apart from the optical bands, we have also looked for the infrared identification of these X-ray sources by searching
the 2MASS catalog (Skrutskie et al. 2006) and the ALLWISE catalog (Wright et al. 2010). 
The results are summarized in Table~4 and Table~5. For computing the probability of chance coincidences, we 
estimated the stellar densities in a $1^{\circ}\times1^{\circ}$ field and find the expected chance coincidences $\lambda$ in 
the X-ray error circles. Except for J1653$\_$X1, the probabilities of one or more optical/IR sources lying in the X-ray error
circles by chance are computed using Equation (1) and are summarized in Table~3,4 and 5. 

For those candidate counterparts having photometric measurements in both optical and infrared regimes, we 
construct their spectral energy distributions (SEDs) and probe their nature in further details. This includes the potential 
counterparts of J1120$\_$X1, J1653$\_$X1, J1729$\_$X2, 
J1729$\_$X4, J1729$\_$X9 and J1946$\_$X1. Since 
2FGL~J1653.6-0159 has its optical and X-ray counterpart (i.e. J1653$\_$X1) securely established and its 
optical spectral and temporal properties have already been explored in details by Kong et al. (2014) and Romani et al. (2014), 
we would not discuss the optical/IR properties of J1653$\_$X1 any further in this work. For the
other five sources under our consideration, we firstly performed the extinction-correction by adopting the 
column densities used in computing their absorption-corrected X-ray fluxes (cf. 
Predehl \& Schmitt 1995; Cardelli et al. 1989). Their de-reddened SEDs were plotted in Figure~\ref{oir_sed}. 

For J1120$\_$X1, we have compared its SED with various stellar spectral models obtained from Pickles (1998). We found that 
it is consistent with that of a K4V star (see Figure~\ref{oir_sed}). 
The SED of J1729$\_$X9 also suggests a thermal origin and is peaked in IR regime.
Since the spectral flux library (i.e. Pickles 1998) that we used in this study only covers
the spectral classes down to M stars, there is no model in this library available for comparing with
the SED of J1729$\_$X9. For constraining its property, we used a simple blackbody model instead.
Since the photometric measurement at 22$\mu$m of J1729$\_$X9 has no error estimate, which is due to the low signal-to-noise
measurement in this band, we discarded it in the analysis. Under this assumption,
the surface temperature of J1729$\_$X9 is found to be of order of
$T\sim3000$~K (see Figure~\ref{oir_sed}) which suggests a possible red dwarf nature.

On the other hand, the SEDs of J1946$\_$X1, J1729$\_$X2 and J1729$\_$X4 do not resemble the thermal emission from a star/blackbody.
Instead, we compare their distributions with a power-law $F_{\nu}\propto\nu^{\alpha}$.
For J1729$\_$X2, the emission from mid-IR to $B$ band can be modeled by a power-law with
$\alpha=-1.24$. Similar behavior has also been found in J1729$\_$X4. As the data point of J1729$\_$X4 at 22$\mu$m has no
error estimate due to its low significance, it was ignored in the analysis. With this measurement excluded, the SED of J1729$\_$X4 can be described
by a power-law with $\alpha=-0.35$. Therefore, these two sources are likely to be AGNs and
can be ruled out as MSP candidates. For J1946$\_$X1, there appears to be an
excess from the best-fit power-law ($\alpha=-1.54$) in the optical regime. Limited by the sparse data, we are not able to
make any firm conclusion about its nature. A dedicated spectroscopic observation is encouraged for J1946$\_$X1.

To search for the temporal variation of these candidate optical/IR counterparts, we 
have looked into the data of {\it Catalina Real-Time Transient Survey} (CRTS) for multi-epoch photometric measurements. 
Ignoring those possible counterparts for 2FGL~J1653.6-0159 (see Kong et al. 2014 for the optical temporal properties of this source), 
Five sources in Table~2 have been observed 
by CRTS for more than one time. Their light curves are shown in Figure~\ref{oir_lc}. Using $\chi^{2}$ test, we determined 
the statistical significance for the temporal variability of their optical emission. The results are summarized in Table~6. 
For J1946$\_$X1, J1729$\_$X2, J1729$\_$X4 and J1729$\_$X12, we do not find any evidence for optical variability. On the other hand,
the optical emission of J1120$\_$X1 is found to be variable at a confidence level of $\sim96\%$. 

As the optical emission of a MSP binary is originated from continuous heating by the radiation 
and relativistic wind outflow from the pulsar, this can possibly result in orbital modulation. As there are more than 200 measurements 
for J1120$\_$X1, we have attempted to search for the periodicity with CRTS data. J1120$\_$X1 was observed by the \emph{Siding Springs Survey} (SSS) 
0.5m Schmidt telescope for 217 exposures and the \emph{Catalina Sky Survey} (CSS) 0.7m Schmidt telescope for 76 exposures. 
Utilizing the online tool of CRTS on the SSS light curve, several periodicity candidates have been revealed. 
The most significant one is found at $P\sim 25.31$ days with a false alarm probability of $1.6\times10^{-5}$. 
The second strong candidate is $P \sim 0.96$ days with a false alarm probability of $1.8 \times 10^{-4}$. 
To cross-check the robustness of these results, we have also performed a more detailed analysis by computing the 
Lomb-Scargle periodogram on the combined SSS+CSS light curve. We have detected a series of peaks that were denoted 
as the harmonics of one-year observational gap. Besides them, two signals are also significantly detected as 
$P \sim 25.36$ days and $P \sim 27.35$ days. To further investigate the effect of observational gaps, 
we examined the power spectrum of the window function. The $\sim$ 1-year and 1-day signals remain the strongest peaks 
in the power spectrum of the window function. Except for these two signals, another strong peak located at $P\sim29.53$ days, 
which corresponds to a synodic month, was also significantly obtained. In addition, several aliases, including $P \sim 27.32$ days 
(a sidereal month) and $P \sim 25.39$ days that representing the beat periods of yearly and monthly signals are also clearly observed. 
Considering the possible differences between SSS and CSS data, we also applied the same analysis on the SSS light curve solely. 
The result is similar to that obtained with the combined light curve although the yearly signal and corresponding harmonics are much weaker. 
Therefore, the putative $\sim 25.36$ days periodicity identified by the online CRTS tool is very likely caused by the non-uniform distribution of data points.

\subsection{Searches for potential radio counterparts}
Since all the known MSPs are radio emitters, we have also searched for possible radio counterparts for all the X-ray sources in Table~2.
Using the point source catalogs of {\it Sydney University Molonglo Sky Survey} (Bock et al. 1999) and 
{\it NRAO VLA Sky Survey} (Condon et al. 1998), we did not find any radio source within $5\sigma$ X-ray positional 
uncertainties of all the tabulated X-ray sources. Barr et al. (2013) have performed a target pulsar survey of 
2FGLs~J1120.0-2204 and J1625.2-0020 with Effelsberg 100~m radio telescope. This survey does not result in any pulsar 
detection associated with these two $\gamma-$ray sources. A limiting flux density of $<70$~$\mu$Jy at 1.36~GHz has been 
placed. 

\section{Summary \& Discussion}
We have systematically searched for MSP candidates from the unidentified 2FGL objects. Seven unassociated 
$\gamma$-ray sources are suggested to be promising candidates. We have also searched for the
X-ray and optical/IR sources within their $\gamma$-ray error ellipses. This enables us to constrain 
the positions
of these $\gamma$-ray selected MSP candidates to a much higher accuracy which facilitates 
the pulsation search and multiwavelength 
follow-up investigations in the future. 

\begin{figure}[t]
\centerline{\psfig{figure=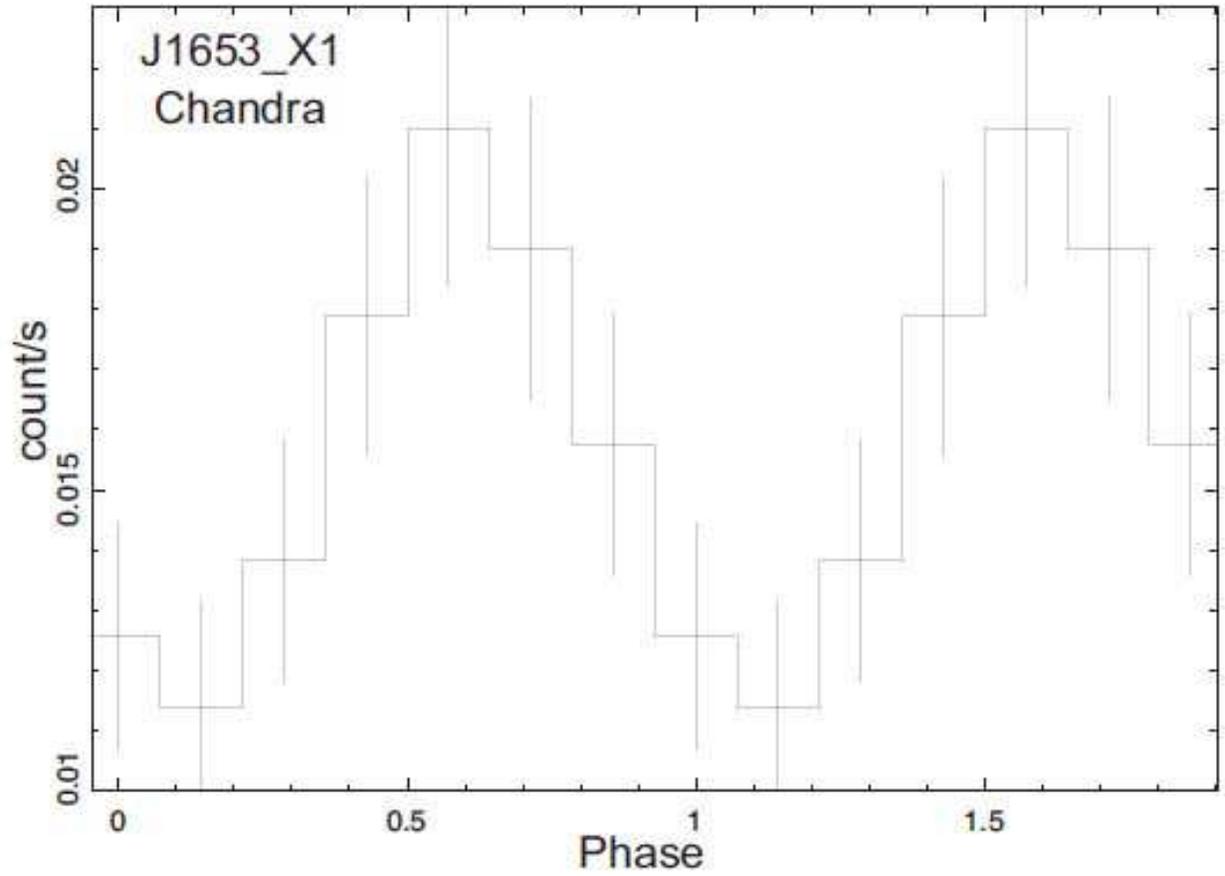,width=17cm,clip=}}
\caption[]{The background-subtracted light curve of J1653$\_$X1 as observed by \emph{Chandra}
in $0.5-8$~keV which is folded at the periodicity of 4448~s. 
The epoch of phase zero is set at the start of the good time interval of this observation.  
Two cycles of orbital modulation are shown for clarity.}
\label{j1653_x1_lc}
\end{figure}

\begin{figure}[t]
\centerline{\psfig{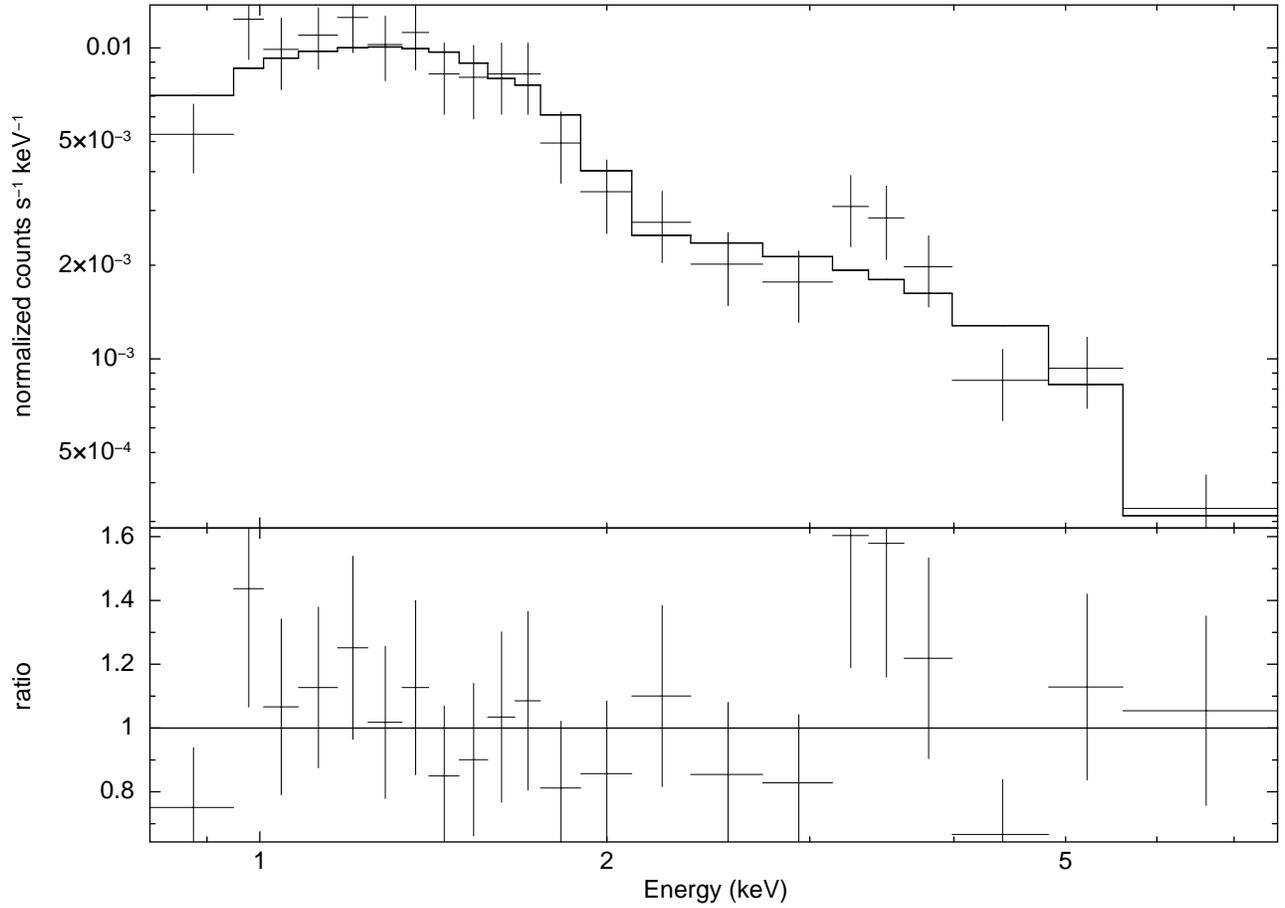}}
\caption[]{The X-ray spectrum of J1653$\_$X1 as observed by \emph{Chandra} and
fitted with an absorbed power-law model (upper panel) and the ratio between 
the observed data and the best-fit model (lower panel). We note the excess at 
energies around 3.5~keV.}
\label{j1653_spec}
\end{figure}

\begin{figure}[t]
\centerline{\psfig{figure=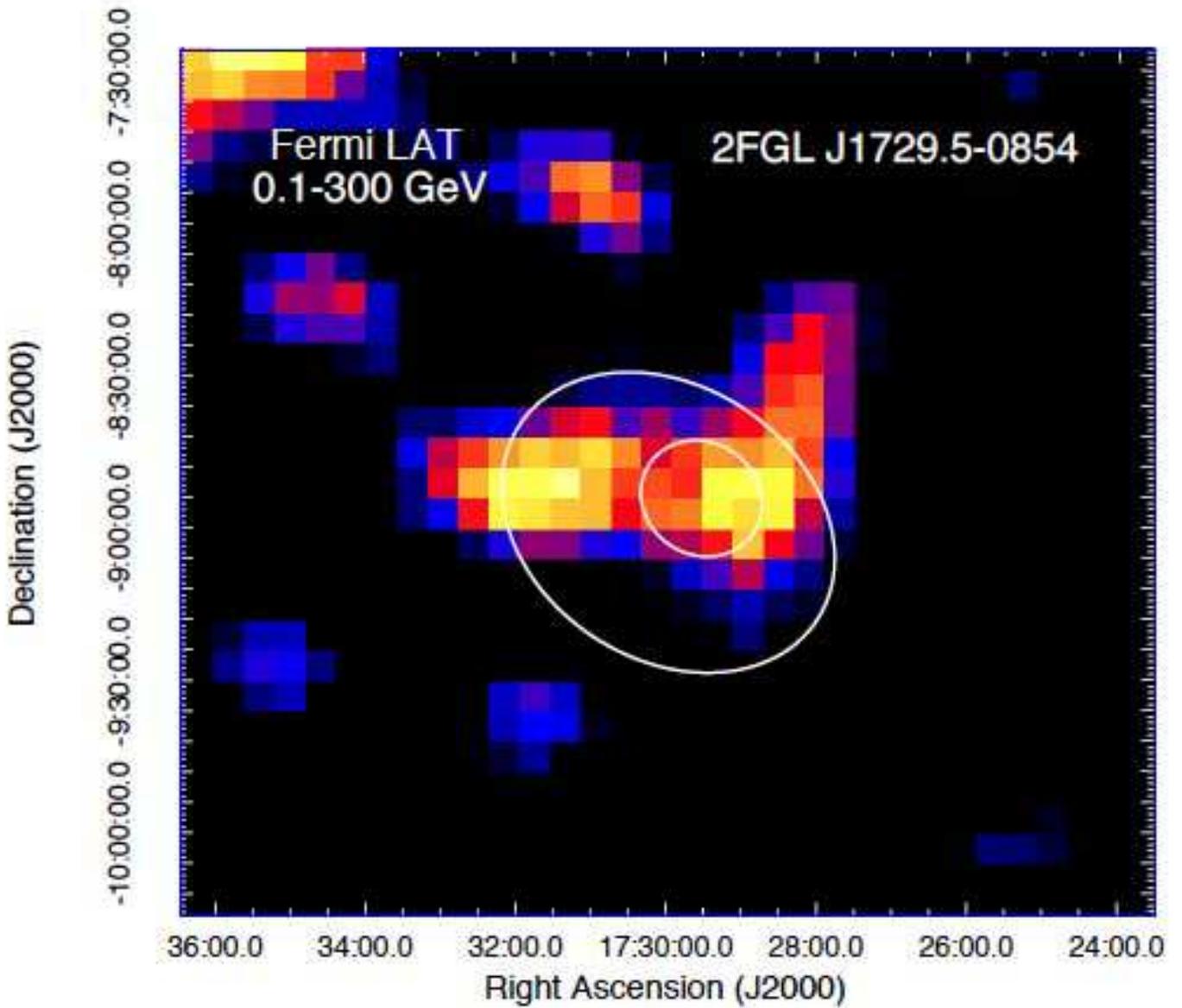,width=19cm,clip=}}
\caption[]{The background-subtracted 0.1--300~GeV $\gamma$-ray count map of 2FGL~J1729.5-0854 as observed by \emph{Fermi}-LAT. 
The image is smoothed with a Gaussian kernal of $\sigma=0.4^{\circ}$. The white circles illustrate the 95\% confidence ellipses in 
2FGL ({\it small}) and 3FGL ({\it large}).}
\label{j1729_lat_img}
\end{figure}

\begin{figure}[t]
\centerline{\psfig{figure=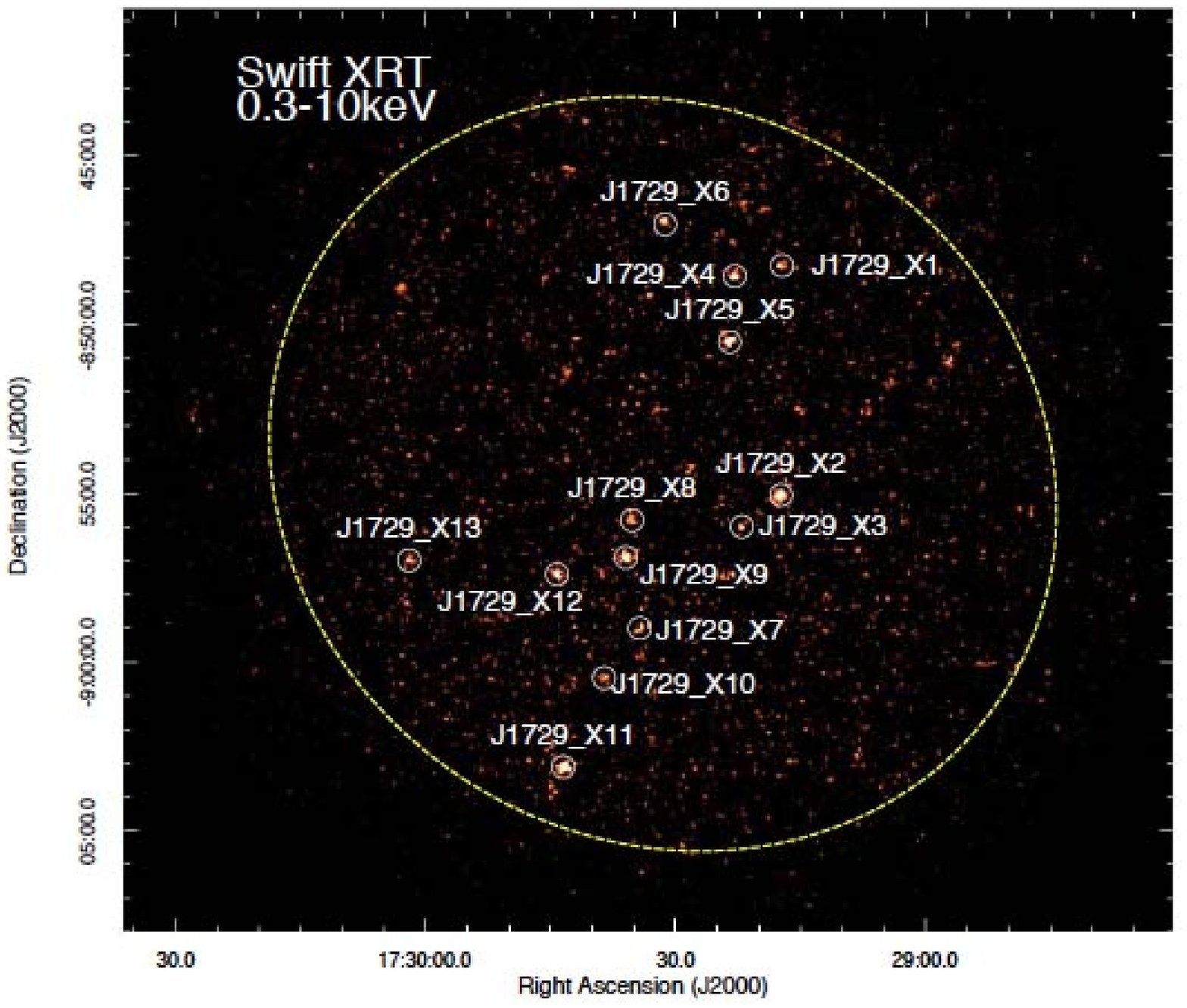,width=19cm,clip=}}
\caption[]{The potential X-ray counterparts (white circles) of 2FGL~J1729.5-0854 as observed by \emph{Swift} XRT. This image is produced by 
using all the available XRT data of this field.
The dashed ellipse illustrates the 2FGL 95\% confidence ellipse. The image is smoothed with a Gaussian kernal of $\sigma=4''$.}
\label{j1729_x_img}
\end{figure}

\begin{figure}[t]
\centerline{\psfig{figure=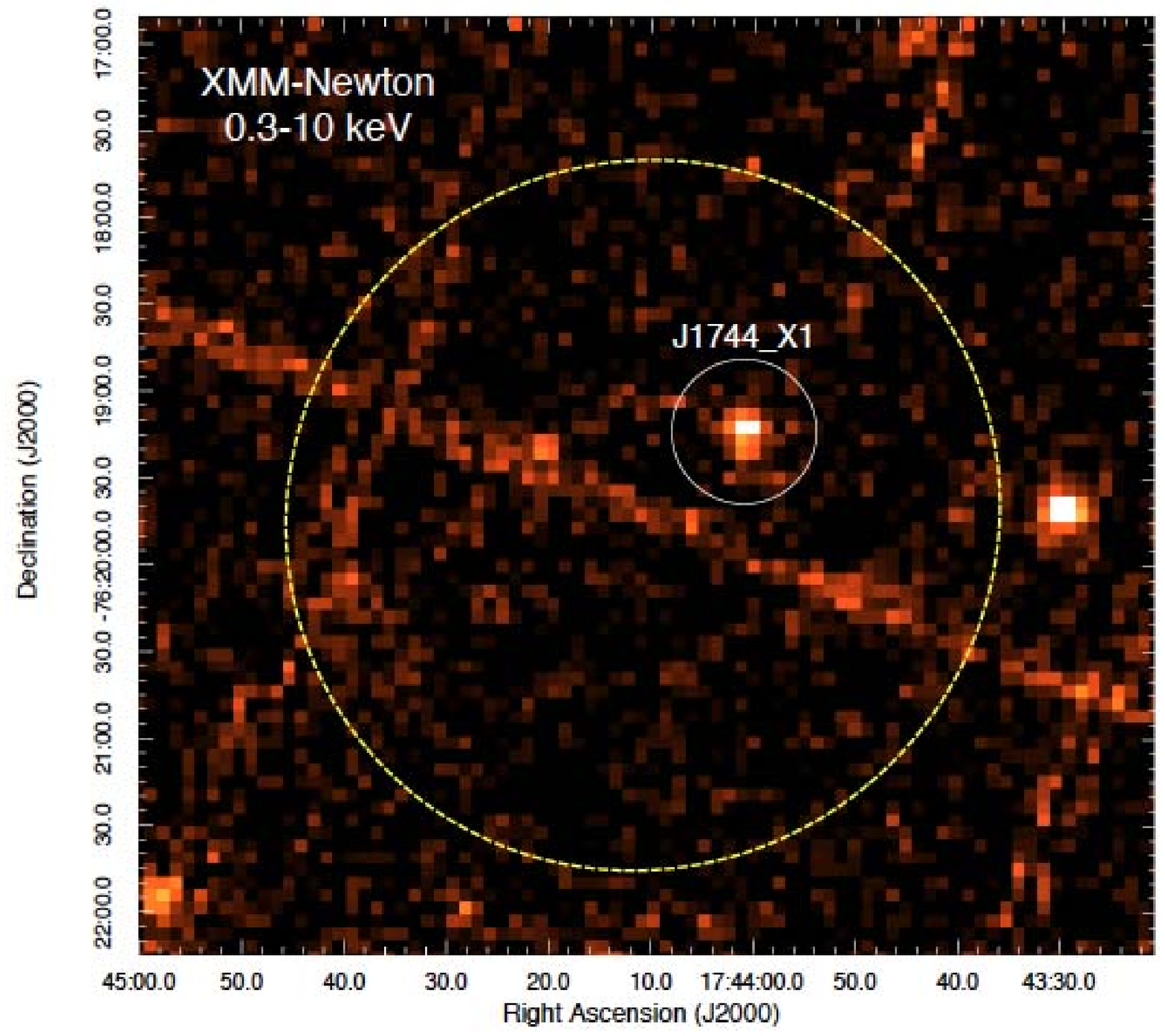,width=19cm,clip=}}
\caption[]{The potential X-ray counterpart (white circle) of 2FGL~J1744.1-7620 as observed by \emph{XMM-Newton}. 
This image is produced by merging all three CCD data. 
The dashed ellipse illustrates the 3FGL 95\% confidence ellipse. The image is smoothed with a Gaussian kernal of $\sigma=4''$.}
\label{j1744_x_img}
\end{figure}

\begin{figure}[t]
\centerline{\psfig{figure=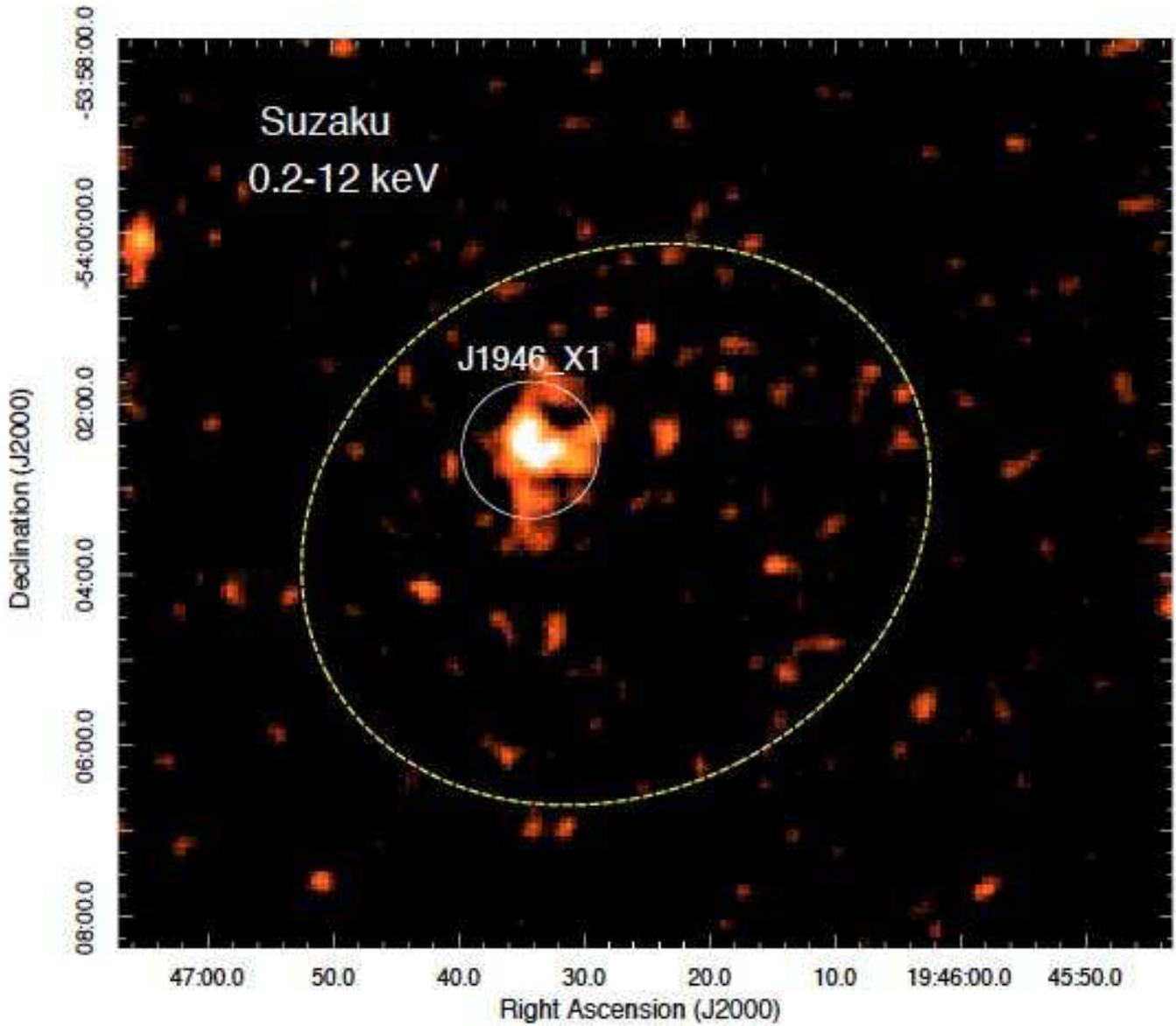,width=19cm,clip=}}
\caption[]{The potential X-ray counterpart (white circle) of 2FGL~J1946.4-5402 as observed by \emph{Suzaku}. This image is produced 
by merging the data from the cameras XIS0, XIS1 and XIS3.
The dashed ellipse illustrates the 3FGL 95\% confidence ellipse. The image is smoothed with a Gaussian kernal of $\sigma=12''$.}
\label{j1946_x_img}
\end{figure}

\begin{figure}[t]
\centerline{\psfig{figure=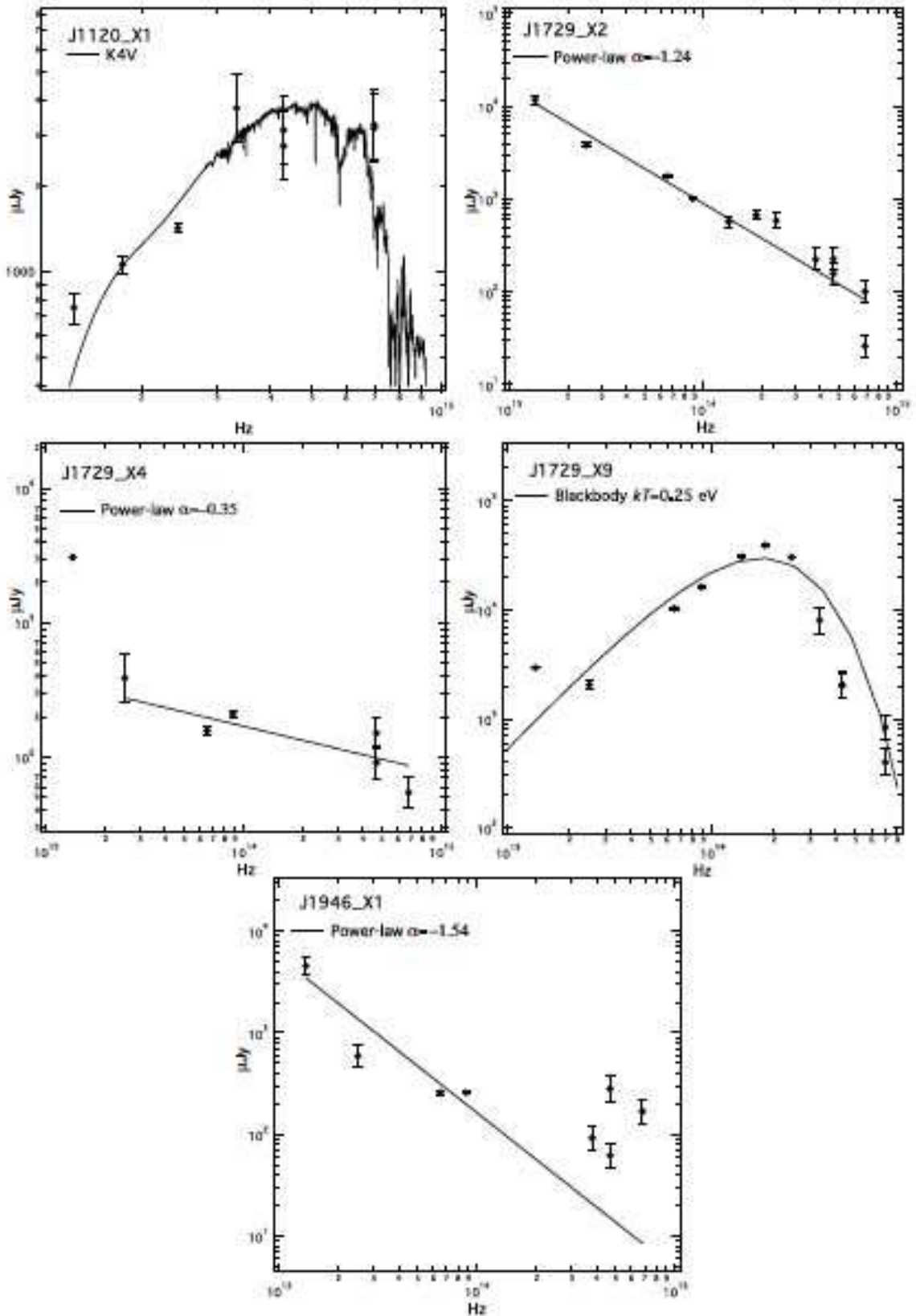,width=16cm,clip=}}
\vspace{-0.5cm}
\caption[]{Spectral energy distributions (SEDs) for the possible optical/infrared counterparts associated with J1120$\_$X1, 
J1729$\_$X2, J1729$\_$X4,J1729$\_$X9, and J1946$\_$X1.
All the data points have been de-reddened. Their distributions 
are compared with a
stellar spectral model (Pickles 1998), a blackbody or a power-law.}
\label{oir_sed}
\end{figure}

\begin{figure}[t]\centerline{\psfig{figure=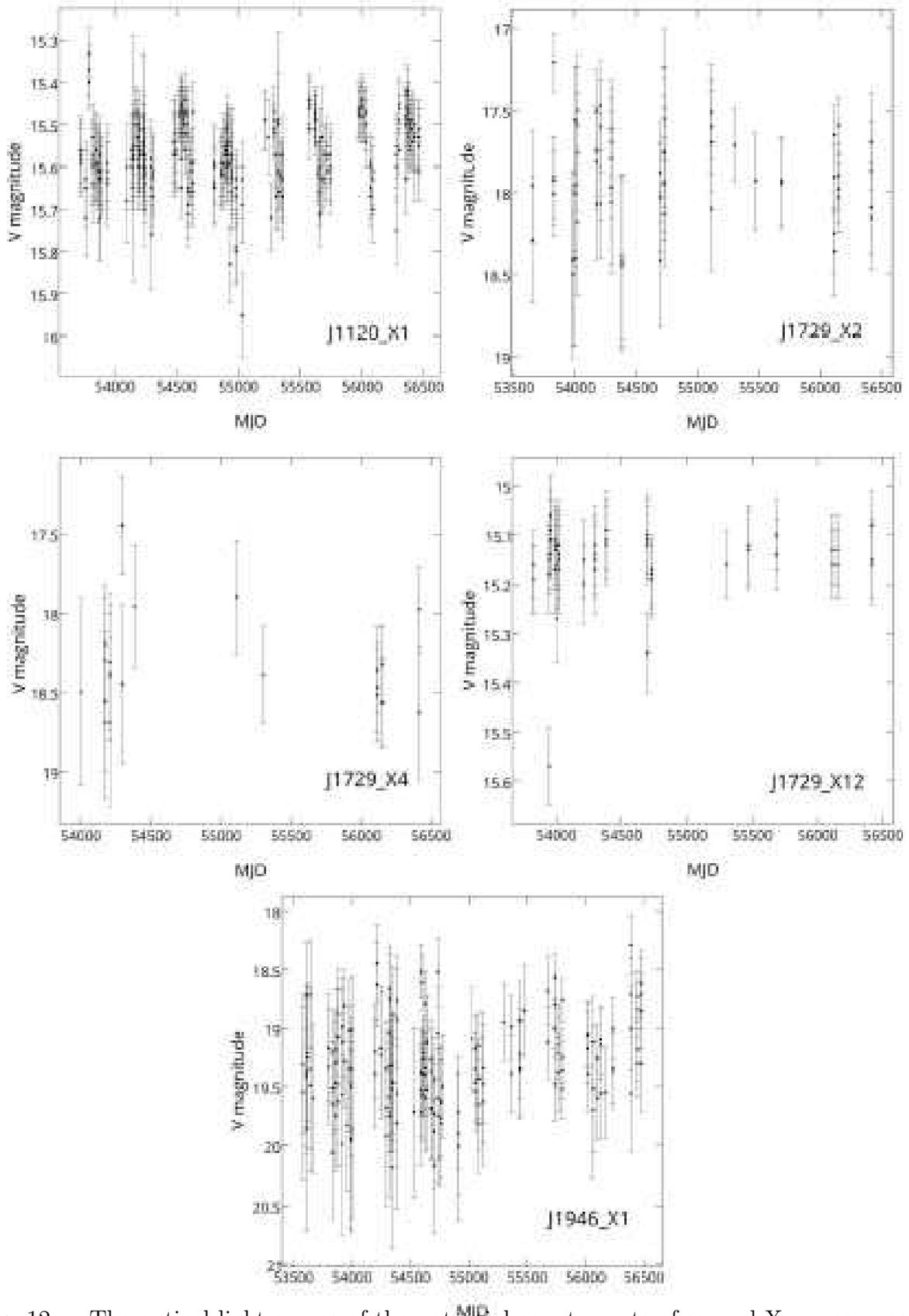,width=16cm,clip=}}
\vspace{-1cm}
\caption[]{The optical light curves of the potential counterparts of several X-ray sources as obtained from CRTS.}
\label{oir_lc}
\end{figure}

Apart from characterizing their X-ray positions and X-ray fluxes through a systematic analysis, 
we have also performed a detailed X-ray spectral analysis for those sources that have sufficient photon 
statistics. In particular, we confirm the non-thermal X-ray nature of J1653$\_$X1 and 
J1946$\_$X1. For a MSP in a tight binary system, the non-thermal X-rays can be resulted from the 
intrabinary shock. The Doppler boosting of the post-shocked pulsar wind can result in the X-ray 
orbital modulation as observed in the case of J1653$\_$X1 (Figure~\ref{j1653_x1_lc}). 
For J1653$\_$X1, besides the power-law spectrum, we have also identified a putative emission feature at $\sim3.5$~keV 
(cf. Figure~\ref{j1653_spec}). 
However, the current data do not allow us to confirm its existence unambiguously. 
Observation with improved photon statistic is needed for a further investigation of this feature. 
Another possible X-ray MSP candidate is J1946$\_$X1. It is the only X-ray source 
found within the $\gamma$-ray error ellipse. The spectral analysis shows that its X-ray emission is also clearly non-thermal dominant. 
Also, the X-ray to $\gamma$-ray flux ratios of J1653$\_$X1 and J1946$\_$X1, $f_{\rm x}/f_{\gamma}\sim10^{-2}$, 
are comparable to the typical values of MSPs or radio-loud pulsar $f_{\rm x}/f_{\gamma}\sim5\times10^{-3}$ 
(Abdo et al. 2013; Marelli et al. 2011). 
This further suggests J1653$\_$X1 and J1946$\_$X1 to be promising pulsar candidates. 

2FGL~J1120.0-2204 is another candidate that deserves a follow-up investigation. 
Through identifying the potential optical/IR counterparts
of the brightest X-ray source J1120$\_$X1 within the $\gamma$-ray error ellipse, 
we have constructed its SED and found that it is comparable with a late-type star. This suggests it to be a  possible 
candidate of a low-mass X-ray binary/redback MSP. Also, its optical light curve as obtained from CRTS 
shows evidence of variability. Therefore, the optical behaviors make J1120$\_$X1 as another interesting 
target for the follow-up studies with dedicated X-ray and optical/IR observations. 

On the other hand, for 2FGL~J1744.1-7620 and 2FGL~J1625.2-0020, we do not identify any possible optical/IR association 
in this study. 
This might indicate they are candidates for isolated MSPs or non-recycled pulsars. Taking a limiting magnitude of 
$V\sim21$ (Monet et al. 2003), the X-ray to optical flux ratios of J1744$\_$X1 and J1625$\_$X1 are found to be 
$f_{\rm x}/f_{\rm opt}\gtrsim10^{-3}$. However, this is not constraining in determining their emission nature 
(e.g. Maccacaro et al. 1988; Stocke et al. 1991). 
Dedicated optical/IR observations at their X-ray positions are required to place tighter constraint. 
We also compare their X-ray to $\gamma-$ray flux ratios with that of non-recycled radio-loud pulsars 
($f_{\rm x}/f_{\rm \gamma}\sim(0.34-53)\times10^{-3}$), 
radio-quiet pulsars ($f_{\rm x}/f_{\rm \gamma}\sim(0.11-1.0)\times10^{-3}$) 
and MSPs ($f_{\rm x}/f_{\rm \gamma}\sim(1.6-15)\times10^{-3}$) 
(Abdo et al. 2013). Within $1\sigma$ uncertainty, the low flux ratio of 2FGL~J1744.1-7620,  
$f_{\rm x}/f_{\rm \gamma}\sim(0.5-0.8)\times10^{-3}$, is consistent with that of 
non-recycled radio-loud/radio-quiet pulsars. On the other hand, the flux ratio of 2FGL~J1625.2-0020, 
$f_{\rm x}/f_{\rm \gamma}\sim(1.0-1.6)\times10^{-3}$, is consistent with all three classes mentioned above. 

The non-detection of any possible radio counterpart for these pulsar candidates can be ascribed to the relatively low sensitvities of 
the existing surveys. 
Targeted observations are encouraged to search for the radio emission from the X-ray/optical positions constrained in this study. 
While radio pulsation search can unambiguously confirm the pulsar nature, radio imaging observation is also useful.
Since some of these MSP candidates can be redbacks, they are subjected to possible state-switch (Takata et al. 2014; Li et al. 2014). 
When they are fliped to accretion-powered state, the increased mass-transfer rate can result in a complicated local ionized environment which 
can smear the pulsed radio emission. In this case, even though the radio pulsar emission mechanism is still active, it will 
be very difficult to detect it. On the other hand, as a direct imaging is free from the problem of correcting complicated 
dispersion, it can be a robust method to investigate if there is any radio emission from the X-ray/optical position. 

\begin{landscape}
\begin{deluxetable}{cccccccc}
\tablewidth{0pc}
\tablecaption{Candidates for millisecond pulsars selected from the unidentified objects in 2FGL catalog.}
\startdata
\hline\hline
Source & RA (J2000) & Dec (J2000) & $b$\tablenotemark{a} & Vari. Index\tablenotemark{b} & Curv. Sig.\tablenotemark{c}  & Detect. Sig.\tablenotemark{d} & $F_{\gamma}$\tablenotemark{e} \\
       & h m s      & d m s       & degree  &           &                &               & 10$^{-11}$ erg/cm$^{2}$/s \\
\hline
2FGL J1120.0-2204   & 11 20 00.6 & -22 04 50  &  36.05  &   25.9   &  5.5    & 20.5 & 1.80$\pm0.15$ \\
2FGL J1539.2-3325   & 15 39 15.1 & -33 25 43  &  17.53  &   29.5   &  5.5    & 10.8 & 1.06$\pm0.15$ \\
2FGL J1625.2-0020   & 16 25 12.8 & -00 20 05  &  31.83  &   24.6   &  8.3    & 20.5 & 1.61$\pm0.13$ \\
2FGL J1653.6-0159   & 16 53 36.6 & -01 59 46  &  24.93  &   17.0   &  5.3    & 22.5 & 3.43$\pm0.25$ \\
2FGL J1729.5-0854   & 17 29 31.5 & -08 54 26  &  13.71  &    9.0   &  7.4    & 10.2 & 1.93$\pm0.20$ \\
2FGL J1744.1-7620   & 17 44 11.0 & -76 20 30  & -22.48  &   27.1   &  5.7    & 20.8 & 2.00$\pm0.19$ \\
2FGL J1946.4-5402   & 19 46 24.4 & -54 02 46  & -29.55 &    24.4   &  5.3    & 12.2 & 0.99$\pm0.13$ \\
\hline
\enddata
\tablenotetext{a}{\footnotesize{Galactic latitude.}}
\tablenotetext{b}{\footnotesize{\texttt{Variability$\_$Index} in 2FGL catalog.}}
\tablenotetext{c}{\footnotesize{\texttt{Curve$\_$Significance} in 2FGL catalog (in unit of $\sigma$).}}
\tablenotetext{d}{\footnotesize{\texttt{Detection$\_$Significance} in 2FGL catalog (in unit of $\sigma$).}}
\tablenotetext{e}{\footnotesize{$\gamma$-ray energy flux in 100~MeV$-$300~GeV.}}
\label{2fgl}
\end{deluxetable}
\end{landscape}

\begin{landscape}
\begin{longtable}{ccccccc}
\caption{Properties of X-ray sources within $\gamma$-ray error ellipses (95\% confidence) of selected 2FGL unidentified objects.}\\
\hline\hline
Source & RA (J2000) & Dec (J2000) & $\sigma_{\rm pos} (a)$ & $F^{\rm obs}_{0.3-10} (b)$ & $F^{\rm unabs}_{0.3-10} (c)$ & $F_{x}/F_{\gamma} (d)$ \\
             & h m s      & d m s       &  arcsec            & erg/cm$^{2}$/s & erg/cm$^{2}$/s &  \\
\hline
\multicolumn{7}{c}{2FGL J1120.0-2204}\\
\hline
J1120$\_$X1 & 11 19 58.376 &  -22 04 55.84 & 0.74 & $(4.1\pm0.5)\times10^{-14}$ & $(4.7\pm0.5)\times10^{-14}$ & $(2.6\pm0.4)\times10^{-3}$ \\
J1120$\_$X2 & 11 20 01.984 &  -22 04 58.65 & 1.34 & $(1.3\pm0.3)\times10^{-14}$ & $(1.5\pm0.3)\times10^{-14}$ & $(8.5\pm2.0)\times10^{-4}$ \\
\hline
\multicolumn{7}{c}{2FGL J1625.2-0020}\\
\hline
J1625$\_$X1 & 16 25 10.469 & -00 21 26.73 & 0.66 & $(1.8\pm0.4)\times10^{-14}$ & $(2.1\pm0.5)\times10^{-14}$ & $(1.3\pm0.3)\times10^{-3}$ \\
\hline
\multicolumn{7}{c}{2FGL J1653.6-0159}\\
\hline
J1653$\_$X1 & 16 53 37.98  & -01 58 37.7 & 0.19 & $(2.4\pm0.1)\times10^{-13}$ & $(3.0\pm0.2)\times10^{-13}$ & $(8.7\pm0.7)\times10^{-3}$ \\
J1653$\_$X2 & 16 53 41.31  & -01 59 27.2 & 1.29 & $(1.7\pm0.4)\times10^{-14}$ & $(2.1\pm0.5)\times10^{-14}$ & $(6.1\pm1.5)\times10^{-4}$ \\
\hline
\multicolumn{7}{c}{2FGL J1729.5-0854}\\
\hline
J1729$\_$X1 & 17 29 17.226 &  -08 48 15.68 & 1.37 & $(1.1\pm0.3)\times10^{-14}$ & $(1.5\pm0.4)\times10^{-14}$ & $(7.9\pm2.3)\times10^{-4}$ \\
J1729$\_$X2 & 17 29 17.352 &  -08 55 03.47 & 0.94 & $(3.9\pm0.6)\times10^{-14}$ & $(5.5\pm0.9)\times10^{-14}$ & $(2.8\pm0.5)\times10^{-3}$ \\
J1729$\_$X3 & 17 29 22.042 &  -08 55 59.67 & 1.20 & $(1.1\pm0.3)\times10^{-14}$ & $(1.5\pm0.4)\times10^{-14}$ & $(7.6\pm2.3)\times10^{-4}$ \\
J1729$\_$X4 & 17 29 22.907 &  -08 48 34.25 & 1.27 & $(2.0\pm0.4)\times10^{-14}$ & $(2.8\pm0.6)\times10^{-14}$ & $(1.5\pm0.3)\times10^{-3}$ \\
J1729$\_$X5 & 17 29 23.507 &  -08 50 31.20 & 1.03 & $(3.1\pm0.5)\times10^{-14}$ & $(4.3\pm0.7)\times10^{-14}$ & $(2.2\pm0.4)\times10^{-3}$ \\
J1729$\_$X6 & 17 29 31.242 &  -08 47 02.19 & 1.57 & $(1.3\pm0.3)\times10^{-14}$ & $(1.7\pm0.5)\times10^{-14}$ & $(9.0\pm2.5)\times10^{-4}$ \\
J1729$\_$X7 & 17 29 34.306 &  -08 59 00.54 & 1.55 & $(1.1\pm0.3)\times10^{-14}$ & $(1.5\pm0.4)\times10^{-14}$ & $(7.9\pm2.3)\times10^{-4}$ \\
J1729$\_$X8 & 17 29 35.205 &  -08 55 48.28 & 1.53 & $(1.1\pm0.3)\times10^{-14}$ & $(1.6\pm0.5)\times10^{-14}$ & $(8.2\pm2.5)\times10^{-4}$ \\
J1729$\_$X9 & 17 29 35.951 &  -08 56 54.59 & 1.09 & $(3.3\pm0.6)\times10^{-14}$ & $(4.6\pm0.8)\times10^{-14}$ & $(2.4\pm0.5)\times10^{-3}$ \\
J1729$\_$X10 & 17 29 38.560 &  -09 00 28.82 & 1.69 & $(1.2\pm0.3)\times10^{-14}$ & $(1.6\pm0.5)\times10^{-14}$ & $(8.5\pm2.5)\times10^{-4}$ \\
J1729$\_$X11 & 17 29 43.483 &  -09 03 07.06 & 1.14 & $(3.0\pm0.5)\times10^{-14}$ & $(4.1\pm0.7)\times10^{-14}$ & $(2.1\pm0.4)\times10^{-3}$ \\
J1729$\_$X12 & 17 29 44.205 &  -08 57 24.41 & 1.19 & $(2.1\pm0.2)\times10^{-14}$ & $(3.0\pm0.3)\times10^{-14}$ & $(1.5\pm0.2)\times10^{-3}$ \\
J1729$\_$X13 & 17 30 01.955 &  -08 56 59.71 & 1.16 & $(1.3\pm0.4)\times10^{-14}$ & $(1.9\pm0.5)\times10^{-14}$ & $(9.6\pm2.7)\times10^{-4}$ \\
\hline
\multicolumn{7}{c}{2FGL J1744.1-7620}\\
\hline
J1744$\_$X1 & 17 44 00.993  & -76 19 13.95 & 0.67 & $(1.0\pm0.2)\times10^{-14}$ & $(1.2\pm0.3)\times10^{-14}$ & $(6.2\pm1.5)\times10^{-4}$  \\
\hline
\multicolumn{7}{c}{2FGL J1946.4-5402}\\
\hline
J1946$\_$X1 & 19 46 33.694 & -54 02 34.91 & 3.9 & $(1.1\pm 0.2)\times10^{-13}$ & $(1.2\pm0.4)\times 10^{-13}$ & $(1.2\pm 0.5)\times10^{-2}$ \\
\hline
\multicolumn{7}{l}{\footnotesize (a) Statistical positional uncertainties.}\\
\multicolumn{7}{l}{\footnotesize (b) Observed fluxes in 0.3-10~keV.}\\
\multicolumn{7}{l}{\footnotesize (c) Absorption-corrected fluxes in 0.3-10~keV.}\\
\multicolumn{7}{l}{\footnotesize (d) X-ray to $\gamma$-ray flux ratios.}\\
\label{msp_x}
\end{longtable}
\end{landscape}

\begin{landscape}
\begin{longtable}{l c c c c c c c c c}
\caption{The potential optical counterparts for the X-ray sources listed in Table~2 as identified in USNO-B1.0 catalog. 
The photometric accuracy is $\sim0.3$~mag.}\\
\hline\hline
Source & RA (J2000) & Dec (J2000) & $B1$ & $R1$ & $B2$ & $R2$ & $I$ & offset $(a)$ & $P(\geq1)$ $(b)$ \\ 
 & d m s & h m s & mag & mag & mag & mag & mag & arcsec & $\%$ \\
\hline
J1120$\_$X1 & 11 19 58.252 & -22 04 56.29 & 15.62 & 15.15 & 15.59 & 15.29 & 14.69 & 1.77 & 3.25 \\
J1653$\_$X1 & 16 53 37.970 & -01 58 36.57 & 19.72 & 19.31 & 20.40 & 19.41 & 20.00 & 1.14 & - \\
J1729$\_$X2 & 17 29 17.280 & -08 55 03.41 & 21.68 & 18.48 & 20.21 & 18.87 & 18.05 & 1.07 & 8.61 \\
J1729$\_$X4 & 17 29 22.923 & -08 48 31.60 & - & 19.15 & 21.29 & 19.70 & - & 2.66 & 9.06 \\
J1729$\_$X9 & 17 29 35.803 & -08 56 53.74 & 18.72 & 16.11 & 17.92 & 16.08 & 14.17 & 2.36 & 8.79 \\
J1946$\_$X1 & 19 46 33.639 & -54 02 36.40 & - & 17.78 & 18.83 & 19.42 & 18.72 & 1.57 & 11.57 \\
\hline
\multicolumn{10}{l}{\footnotesize (a) Angular distance between the X-ray positions and the proper-motion corrected optical positions.}\\
\multicolumn{10}{l}{\footnotesize (b) Probability of one or more optical sources lying in the X-ray error circles by chance.}\\
\end{longtable}
\end{landscape}

\begin{landscape}
\begin{longtable}{c c c c c c c c}
\caption{The potential infrared counterparts for the X-ray sources listed in Table~2 as identified in 2MASS catalog.}\\
\hline\hline
Source & RA (J2000) & Dec (J2000) & $ J $ & $ H $ & $ K_s $ & offset $(a)$ & $P(\geq1)$ $(b)$ \\
 & h m s & d m s & mag & mag & mag & arcsec & $\%$ \\
\hline
J1120$\_$X1 & 11 19 58.356 & -22 04 56.65 & $15.19\pm0.04 $ & $15.06\pm0.08$ & $14.90\pm0.14$ & 0.86 & 0.90 \\
J1729$\_$X2 & 17 29 17.302 & -08 55 03.57 & $16.33\pm0.11 $ & $15.68\pm0.11$ & $15.27\pm0.15$ & 0.75 & 6.11 \\
J1729$\_$X9 & 17 29 35.801 & -08 56 53.17 & $12.05\pm0.02 $ & $11.27\pm0.02$ & $10.95\pm0.02$ & 2.65 & 6.29 \\
\hline
\multicolumn{8}{l}{\footnotesize (a) Angular distance between the X-ray positions and the IR positions.}\\
\multicolumn{8}{l}{\footnotesize (b) Probability of one or more infrared sources lying in the X-ray error circles by chance.}\\
\end{longtable}
\end{landscape}

\begin{landscape}
\begin{longtable}{l c c c c c c c c}
\caption{The potential infrared counterparts for the X-ray sources listed in Table~2 as identified in WISE catalog.}\\
\hline\hline
Source & RA (J2000) & Dec (J2000) & $W1$ ($3.4\mu$m) & $W2$ ($4.6\mu$m)  & $W3$ ($12\mu$m)  & $W4$ ($22\mu$m)  & offset $(a)$  & $P(\geq1)$ $(b)$ \\
 & h m s & d m s & mag & mag & mag & mag & arcsec & $\%$ \\
\hline
J1120$\_$X2 & 11 20 01.787 & -22 04 57.04 & $15.64\pm0.04$ & $15.10\pm0.08$ & $11.68\pm0.25$ & $8.70$ & 3.14 & 4.78 \\
J1729$\_$X2 & 17 29 17.316 & -08 55 03.66 & $13.79\pm0.03$ & $12.52\pm0.02$ & $9.80\pm0.06$ & $7.13\pm0.11$ & 0.53 & 7.32 \\
J1729$\_$X4 & 17 29 22.933 & -08 48 32.80 & $15.50\pm0.050$ & $15.14\pm0.09$ & $12.28\pm0.43$ & $8.59$ & 1.47 & 7.69 \\
J1729$\_$X9 & 17 29 35.796 & -08 56 53.42 & $10.78\pm0.02$ & $10.60\pm0.02$ & $10.47\pm0.08$ & $8.64$ & 2.56 & 7.50 \\
J1729$\_$X10 & 17 29 38.532 & -09 00 25.54 & $16.66\pm0.11$ & $16.44$ & $12.38$ & $8.99$ & 3.32 & 8.42\\
J1729$\_$X12 & 17 29 43.975 & -08 57 24.14 & $11.21\pm0.02$ & $11.02\pm0.02$ & $10.93\pm0.15$ & $8.94$ & 3.50 & 7.60\\
J1729$\_$X13 & 17 30 01.947 & -08 56 57.14 & $15.93\pm0.06$ & $15.00\pm0.08$ & $12.27\pm0.47$ & $8.14$ & 2.58 & 7.60 \\
J1946$\_$X1 & 19 46 33.630 & -54 02 36.43 & $15.20\pm0.04$ & $14.56\pm0.05$ & $11.83\pm0.26$ & $8.16\pm0.23$ & 1.62 & 12.28 \\
\hline
\multicolumn{8}{l}{\footnotesize (a) Angular distance between the X-ray positions and the IR positions.}\\
\multicolumn{8}{l}{\footnotesize (b) Probability of one or more infrared sources lying in the X-ray error circles by chance.}\\
\end{longtable}
\end{landscape}

\begin{longtable}{l c c c}
\caption{Statistical significances for the variabilities of the CRTS optical light curves (Figure~\ref{oir_lc}) 
as determined by $\chi^{2}$ test.}\\
\hline\hline
Source & $\chi^{2}$ & d.o.f. & $p$-value (right tail)\\
\hline
J1120$\_$X1 & 337.53 & 292 & 0.034 \\
J1729$\_$X2 & 16.32 & 22 & 0.80 \\
J1729$\_$X4 & 16.32 & 22 & 0.80 \\
J1729$\_$X12 & 45.93 & 57 & 0.85 \\
J1946$\_$X1 & 166.82  & 167  & 0.49 \\
\hline
\end{longtable}

\acknowledgments{\small
CYH is supported by the National Research Foundation of Korea through grant 2014R1A1A2058590. 
SMP is supported by BK21 plus program. 
CPH is supported by the Ministry of Science and Technology of Taiwan through the grant 102-2112-M-008-020-MY3 and NSC 101-2119-M-008-007-MY3.
AKHK, KLL, RJ and TCY are supported by the Ministry of Science and Technology of Taiwan
through grants 100-2628-M-007-002-MY3, 100-2923-M-007-001-MY3 and 103-2628-M-007-003-MY3.
PHT is supported by the One Hundred Talents Program of the Sun Yat-Sen University.
JT and KSC are supported by a 2014 GRF grant of Hong Kong Government under HKU 17300814P. 
We thank the \emph{Swift} team for scheduling the ToO observations.
We also thank Dr. Xian Hou for reading through the manuscript carefully and provided us with comments 
for improving the quality of this work.
}

\end{document}